\documentclass[11pt]{article}
\usepackage{amsmath,amssymb}
\usepackage{url}
\usepackage{citesort}
\usepackage{latexsym}
\usepackage{amsfonts}
\usepackage{epsfig}
\usepackage{picinpar}
\usepackage{fullpage}

\newtheorem{example}{Example}
\newtheorem{theorem}{Theorem} 
\newtheorem{proposition}{Proposition}
\newtheorem{lemma}{Lemma}
 
\newtheorem{definition}{Definition}
\newtheorem{remark}{Remark} 

\newtheorem{corollary}{Corollary}[section]
\newtheorem{observation}{Observation}[section]

\date{}

\newcommand{\CASP}{$\mathbb{ASP}^A$}

\def\qed{\hfill{$\Box$}}
\def\naf{ not \;}
\def\cal{\mathcal}

\begin{document}

\title{An Unfolding-Based Semantics for Logic Programming with Aggregates} 

\author{Tran Cao Son, Enrico Pontelli, Islam Elkabani \\
Computer Science Department \\
New Mexico State University \\
Las Cruces, NM 88003, USA\\
\{tson,epontell,ielkaban\}@cs.nmsu.edu
}

\maketitle

\begin{abstract}
The paper presents two equivalent definitions of answer sets 
for logic programs with aggregates. These definitions build
on the notion of \emph{unfolding of aggregates}, and they are aimed at
creating methodologies to translate logic programs with aggregates
to normal logic programs or positive programs, whose answer set semantics
can be used to defined the semantics of the original programs. 

The first definition provides an alternative view of the semantics
for logic programming 
with aggregates described in~\cite{PelovDB04,Pelov04}. In particular, 
the unfolding employed by the first definition in this paper coincides with the 
translation of programs with aggregates into normal logic programs 
described in \cite{PelovDB03}. This indicates that the approach proposed in this 
paper captures the same meaning as the semantics discussed in \cite{PelovDB04,Pelov04}.

The second definition is similar to the traditional answer set definition 
for normal logic programs, in that, given a logic program with
aggregates and an interpretation,  the unfolding process produces a
positive program. The paper shows how this definition can be extended to 
consider aggregates in the head of the rules.  

These two approaches are very intuitive, general, 
and do not impose any syntactic restrictions on the
use of aggregates, including support for use of aggregates as heads of
program rules.
The proposed views of logic programming with aggregates
are simple and coincide with the ultimate stable model semantics 
\cite{PelovDB04,Pelov04}, and with other semantic characterizations for
large classes of program (e.g., programs with monotone aggregates
and programs that are aggregate-stratified). 
Moreover, it can be directly employed to support an implementation using available 
answer set solvers. The paper 
describes a system, called \CASP, 
that is capable of computing answer sets of 
programs with arbitrary (e.g., recursively defined) aggregates. 
The paper also presents an experimental  comparison of \CASP\  
with another system for computing answer sets of 
programs with aggregates,  {\tt DLV$^{\mathcal{A}}$}. 
\end{abstract}

\newpage

{\footnotesize
\tableofcontents
}

\newpage

\section{Background and Motivation}
The handling  of 
aggregates in Logic Programming (LP) has been the subject of 
intense studies
in the late 80's and early 90's~\cite{KempS91,MumickPR90,RossS97,van91,ZanioloAO93}. 
Most of these proposals 
focused  on the theoretical foundations  and
computational  properties of aggregate functions 
in LP. The recent development of the {\em answer set 
programming} paradigm, whose underlying 
theoretical foundation is the answer set semantics \cite{GL88},
has renewed the interest in the 
treatment of aggregates 
in LP, and led to a number of new proposals 
\cite{armi-ijcai03,DeneckerPB01,ElkabaniPS04,faberLP04,ferraris,a-prolog,trus,PelovDB04,PelovDB03,Pelov04,SonP05}. 
Unlike many of the earlier proposals, these new efforts provide
a sensible semantics for programs that makes a general 
use of aggregates, including the presence of
\emph{recursion} through the aggregates and the ability to 
use non-monotone aggregate functions.
Most of 
these new efforts build on the spirit of answer set semantics for
LP, and some have found their way in concrete 
implementations. For example,
the current release 
(built BEN/Jan 13 2006)\footnote{\url{http://www.dbai.tuwien.ac.at/proj/dlv}} 
of {\tt DLV$^{\mathcal{A}}$} 
handles aggregate-stratified programs \cite{armi-ijcai03}, and 
the system described in \cite{ElkabaniPS04} supports recursive
aggregates according to the semantics described in~\cite{KempS91}.
A prototype of the  ASET-Prolog system, capable of supporting recursive aggregates,
has also been developed~\cite{mary}.

\smallskip

Answer set semantics 
for LP \cite{GL88} has been one of the most widely adopted semantics for 
\emph{normal} logic programs---i.e., logic programs that allow negation as
failure in the body of the rules.
It is a natural extension of the minimal model 
semantics of  positive logic programs to the case of normal logic programs. 
Answer set semantics provides the theoretical foundation for the recently 
emerging programming paradigm called {\em answer set programming} \cite{Lifschitz02,mar99,nie99}
which has proved to be useful in several applications \cite{Baral03,tplp03-special,Lifschitz02}.

A set of atoms $S$ is an \emph{answer set} of the program $P$ if $S$ is 
the minimal model of the positive program $P^S$ (the \emph{reduct of $P$ 
with respect to $S$}), obtained by 
\begin{itemize}
\item[(i)] removing 
from $P$ all the rules whose body contains a negation as failure
literal $not\:b$ which is false in $S$ (i.e., $b \in S$);
and 
\item[(ii)] removing all the negation as failure literals from the remaining rules. 
\end{itemize}
The above transformation is often referred to as the 
\emph{Gelfond-Lifschitz transformation}. 

This definition of answer sets satisfies 
several important properties. In particular, answer sets are 
\begin{itemize}
\item[] $\mathbf{(Pr_1)}$ {\em closed}, 
	i.e., if an answer set satisfies the body of a rule 
	$r$ then it also satisfies its head; 
\item[] $\mathbf{(Pr_2)}$ 
	 {\em supported}---i.e., for each member $p$ of an answer set $S$
	there exists a rule $r \in P$ such that  $p$ is the head of the
	rule and the body of $r$ is true in $S$;
\item[] $\mathbf{(Pr_3)}$ {\em minimal}---i.e., no proper subset of an 
	answer set is also an answer set. 
\end{itemize}
It should be emphasized that the properties
$\mathbf{(Pr_1)}$-$\mathbf{(Pr_3)}$  are necessary but not sufficient conditions
for a set $S$ to be an answer set of a program $P$.
For example, 
the set $\{p\}$ is not an answer set of the program 
$\{p \leftarrow p, \:\:\: q \leftarrow \naf p\}$, 
even though it satisfies the three properties. Nevertheless, 
these properties constitute the main 
principles that guided several extensions of the answer set semantics 
to different classes of logic programs, such as 
\emph{extended} and
 \emph{disjunctive} logic programs \cite{gel91b}, programs with 
\emph{weight constraint} rules \cite{weight}, and programs with 
\emph{aggregates} (e.g., \cite{armi-ijcai03,KempS91}).
It should also be  mentioned that, for certain classes of logic programs 
(e.g., programs with weight constraints and choice rules \cite{weight}  
or with nested expressions \cite{LifschitzTT99}), 
$\mathbf{(Pr_3)}$ is not satisfied. It is, however, generally accepted 
that $\mathbf{(Pr_1)}$ and $\mathbf{(Pr_2)}$ must be satisfied by any 
answer set definition for any extension of logic programs. 

\smallskip
As  evident from the literature,  a straightforward extension of 
the Gelfond-Lifschitz transformation to programs with 
aggregates leads to the loss of some of the properties
$\mathbf{(Pr_1)}$-$\mathbf{(Pr_3)}$ (e.g., presence of non-minimal answer sets
\cite{KempS91}). Sufficient conditions, that characterize classes of programs
with aggregates for which the properties $\mathbf{(Pr_1)}$-$\mathbf{(Pr_3)}$ of answer
sets hold, have been 
investigated, such as aggregate-stratification and monotonicity
(e.g., \cite{MumickPR90}).
Alternatively, researchers have either accepted the loss
of some of the properties $\mathbf{(Pr_1)}$-$\mathbf{(Pr_3)}$  (e.g., acceptance of
non-minimal answer sets \cite{ElkabaniPS04,a-prolog,KempS91}) or
have \emph{explicitly} introduced  minimality or 
analogous properties as requirements in the definition of answer sets
for programs with aggregates (e.g., \cite{faberLP04,ferraris}). 

\smallskip

The various approaches for
 defining answer set semantics for logic programs with arbitrary 
aggregates differ from each other in both the languages that are considered and 
in the treatment of aggregates. Some proposals accept languages in which
aggregates, or atoms representing aggregates 
(e.g., the weight constraints in {\sc Smodels}-notation), are allowed to occur 
in the head of programs' rules or as facts 
in \cite{ferraris,trus,weight}, while this has been disallowed in other proposals
\cite{armi-ijcai03,DeneckerPB01,ElkabaniPS04,faberLP04,a-prolog,PelovDB04,Pelov04}.
 The advantage of allowing aggregates in the head 
can be seen in the use of choice rules and weight constraints in generate and 
test programs. Allowing aggregates in the head can make the 
encoding of a problem significantly more declarative and compact. Similarly,
some proposals do not consider negation-as-failure literals with aggregates 
\cite{ElkabaniPS04,a-prolog}.

The recent approaches for defining answer sets for logic programs with arbitrary 
aggregates can be roughly divided  into three different groups. The first group can be viewed 
as a straightforward generalization of the work in \cite{GL88}, by treating 
aggregates in the same way as negation-as-failure literals. Belonging to this group 
are the proposals in \cite{ElkabaniPS04,a-prolog,KempS91}. A limitation of this approach is 
that it leads to the acceptance of unintuitive answer sets, in presence of recursion through 
aggregates. Another line of work is to replace aggregates with equivalent 
formulae, according to some notion of equivalence, and to reduce programs 
with aggregates to programs for which the semantics has already been defined 
\cite{ElkabaniPS04,ferraris,PelovDB03}. A third direction is to make use of 
novel semantic constructions 
\cite{DeneckerPB01,PelovDB04,Pelov04,faberLP04,trus,SonP05}.

\smallskip

The objective of this paper is to investigate an alternative
characterization of the semantics of logic programs with
unrestricted use of aggregates. In this context, aggregates are
simply viewed as a syntactic sugar, representing a collection of 
constraints on the admissible interpretations.  
The proposed characterization is designed  to 
maintain the positive properties of the most recent
proposals developed to address this problem (e.g.,
\cite{faberLP04,ferraris,Pelov04}), and to meet the following
requirements:
\begin{itemize}
\item It should apply to programs with \emph{arbitrary} aggregates
	(e.g., no syntactic restrictions in the use of aggregates
	as well as no restrictions on the types of aggregates that
	can be used). In particular, we wish the approach to naturally
	support aggregates as facts and as heads of rules.
\item  It should be as intuitive as the traditional 
	 answer set semantics, and it should extend traditional 
	answer set semantics---i.e., it should behave as traditional
	answer set semantics for programs without aggregates.
	 It should also naturally satisfy the basic properties 
	$\mathbf{(Pr_1)}$-$\mathbf{(Pr_3)}$ 
	of answer sets.

\item  It should offer  ways to implement the semantic
	characterization by integrating, with minimal 
	modifications, the definition 
	in state-of-the-art answer set solvers,
	such as {\sc Smodels} \cite{simons97}, 
	{\sc dlv} \cite{eiter98a}, {\sc Cmodels} \cite{cmodels2},
	{\sc ASSAT} \cite{lin02a}, etc. 
	In particular,
	it should  require little more than the addition of a module to determine
	the \emph{``solutions''} of an aggregate,\footnote{This concept
	is formalized later in the paper.} without
	substantial  modifications of the mechanisms to compute answer sets.

\end{itemize}
We achieve these objectives by defining a transformation, 
called {\em unfolding}, from logic programs with aggregates 
to normal logic programs. The key idea that makes this possible 
is the generalization of the supportedness property of answer sets 
to the case of aggregates. More precisely, our transformation ensures 
that, if an aggregate atom is satisfied by a model 
$M$\/, then $M$ supports at least one of its solutions. 
Solutions of aggregates can be precomputed, and 
an answer set solver for LP with aggregates
can be implemented using standard answer set solvers. 

The notion of unfolding has been widely used in various
areas of logic programming (e.g.,~\cite{abhik,pettorossi,satotamaki}). The 
inspiration for the approach used in handling aggregates in
this paper comes from the methodology proposed in various works
on \emph{constructive negation} (e.g., \cite{stuckey1,chan,neg})---in particular,
from the idea of unfolding \emph{intensional sets} into sets of 
solutions, employed to handle intensional sets in~\cite{sets94,neg}. 

The approach developed in this paper is the continuation 
and improvement of the approach in \cite{ElkabaniPS04}. 
It offers an alternative view of the semantics for LP with aggregates 
developed in~\cite{Pelov04}. In particular, the two characterizations
provide the same meaning to program with aggregates, although our approach does
not require the use of  approximation theory. 
We provide two ways of using unfolding. The first is similar to the 
notion of transformation explored in \cite{PelovDB03}. The second is 
closer to the spirit of the original definition of answer sets \cite{GL88}, 
and it allows us to naturally handle more general use of aggregates (e.g.,
aggregates in the heads). 
The characterization proposed in this paper also
captures the same meaning as the proposals 
in~\cite{faberLP04,ferraris,trus} for large classes of programs
(e.g., stratified programs and programs with monotone aggregates). 
Observe that, in this work, we do not directly address the problem of negated 
aggregates. This problem can be tackled in different ways 
(e.g. \cite{trus,ferraris}). Our approach to aggregates  can 
be easily extended to accommodate any of these approaches \cite{SonPT06}. 

\bigskip

The rest of this paper is organized as follows. Section~\ref{sec2} presents
the syntax of our logic programming language with aggregates.
Section~\ref{semantics} describes the first definition of answer sets 
for programs with aggregates that  do not allow for aggregates to occur 
in the head of rules. The definition is based on an unfolding transformation 
of programs with aggregates into normal logic programs. 
It also contains a discussion of properties of answer sets and 
describes an implementation.
Section~\ref{sec-alternative} introduces an alternative unfolding, 
which is useful for extending the use of aggregates to
the head of program rules. 
Section~\ref{related-work} compares
our approach with the relevant literature. Section~\ref{sec-diss} discusses
some issues related to our approach to providing  semantics of aggregates.
Finally, Section~\ref{concl}
presents the  conclusions and the future work.

\section{A Logic Programming Language with Aggregates} 
\label{sec2}\label{syntax}


Let us consider a signature 
$\Sigma_L = \langle {\cal F}_L\cup {\cal F}_{Agg}, {\cal V}\cup {\cal V}_l, \Pi_L \cup \Pi_{Agg} \rangle$, 
where 
\begin{itemize}
\item ${\cal F}_L$ is a 
collection of constants (\emph{program constants}), 
\item ${\cal F}_{Agg}$ is a collection of unary function
	symbols (\emph{aggregate functions}),
\item ${\cal V}$ and ${\cal V}_l$ are  denumerable collections of variables,
	such that ${\cal V} \cap {\cal V}_l = \emptyset$, 
\item $\Pi_L$ is a collection of arbitrary predicate symbols (\emph{program predicates}), and
\item $\Pi_{Agg}$ is a collection of unary predicate symbols (\emph{aggregate predicates}).
\end{itemize}
In the rest of this paper, we 
will  assume that $\mathbb{Z}$ is 
a subset of ${\cal F}_L$---i.e., there are distinct constants representing the 
integer numbers. 
We will refer to  $\Sigma_L$ as the \emph{ASP signature}.

We will also refer to 
$\Sigma_P = \langle {\cal F}_P,{\cal V}\cup {\cal V}_l, \Pi_P \rangle$
as the \emph{program signature},
where 
\begin{itemize}
\item ${\cal F}_P \subseteq {\cal F}_L$,  
\item $\Pi_P \subseteq \Pi_L$, and
\item ${\cal F}_P$ is finite.
\end{itemize}
We will denote 
with ${\cal H}_P$ the $\Sigma_P$-Herbrand universe, containing
the ground terms built using symbols 
of ${\cal F}_P$\/, and
with ${\cal B}_P$ the corresponding $\Sigma_P$-Herbrand base. 
We will refer to an atom of the form 
$p(t_1,\dots,t_n)$, where $t_i \in {\cal F}_P \cup {\cal V}$ 
and $p\in \Pi_P$, as an ASP-atom. An ASP-literal is either an ASP-atom 
or the negation as failure ($not\:A$) of an ASP-atom. 

\begin{definition}\label{groundset}
An \emph{extensional set} has the form $\{t_1,\dots,t_k\}$, 
where $t_i$ are terms of $\Sigma_P$. An \emph{extensional multiset} 
has  the form 
$\{\!\!\{t_1,\dots,t_k\}\!\!\}$ where $t_i$ are (possibly repeated) terms
of $\Sigma_P$. 
\end{definition}

\begin{definition}\label{intensionalset}
An \emph{intensional set} is of the form 
$$ \{X \:\mid\: p(X_1,\dots,X_k)\} $$
where $X\in {\cal V}_l$ is a variable, $X_i$'s are variables or constants,
$\{X_1,\dots,X_k\} \cap {\cal V}_l = \{X\}$,
and $p$ is a k-ary predicate in $\Pi_P$.

An \emph{intensional multiset} is 
of the form 
$$
\{\!\!\{ X \:\mid\:  \exists Z_1,\dots,Z_r.\: 
p(Y_1,\dots, Y_m)\}\!\!\}
$$
where $\{Z_1,\dots,Z_r,X\}\subseteq {\cal V}_l$,
$Y_1,\dots,Y_m$ are variables or constants
(of ${\cal F}_P$),  $\{Y_1,\dots,Y_m\}\cap {\cal V}_l=\{X,Z_1,\dots,Z_r\}$,
and $X \notin \{Z_1,\dots,Z_r\}$. 
We call $X$ and $p$ 
the \emph{collected variable} and 
the {\em predicate} of the set/multiset, respectively.
\end{definition}
Intuitively, we are collecting 
the values of $X$ that satisfy the atom $p(Y_1,\dots, Y_m)$, under the 
assumption that the variables $Z_j$ are locally and existentially quantified. 
For example, if $p(X,Z)$ is true for $X=1, Z=2$ and $X=1, Z=3$, then the 
multiset $\{\!\!\{X \:\mid\:\exists Z . p(X,Z)\}\!\!\}$ corresponds to 
$\{\!\!\{1,1\}\!\!\}$.
Definition \ref{intensionalset} can be extended 
to allow more complex types of sets, e.g., sets collecting
tuples as elements,  sets with conjunctions of literals
as property of the intensional construction, and
intensional sets with existentially quantified variables. 

Observe also that the variables from ${\cal V}_l$ are used exclusively
as collected or local variables in defining intensional sets or multisets,
and they cannot occur anywhere else.
\begin{definition}
An \emph{aggregate term} is of the form $f(s)$, where $s$ is an
intensional set or multiset, and $f \in {\cal F}_{Aggr}$.
An \emph{aggregate atom} has the form 
$p(\alpha)$
where $p \in \Pi_{Agg}$ and $\alpha$ is an aggregate term.
\end{definition}
This notation for aggregate atoms is more general
than the one used in some previous works, and resembles
the \emph{abstract constraint atom} notation presented
in~\cite{trus}. 

In our examples, we will focus on the
``standard'' aggregate functions and predicates, e.g.,
{\sc Count, Sum, Min, Max, Avg} applied to sets/multisets
and predicates such as $=$, $\neq$, $\leq$, etc. Also, for the
sake of readability, we will often use a more traditional 
notation when dealing with the standard aggregates; e.g.,
instead of writing $\leq_7(\textsc{Sum}(\{X\:|\:p(X)\}))$ we will
use the more common format $\textsc{Sum}(\{X\:|\:p(X)\})\leq 7$.

Given an aggregate atom $\ell$, with k-ary collected predicate $p$, we denote
with ${\cal H}(\ell)$ the following set of ASP-atoms:
\[ {\cal H}(\ell) = \{p(a_1,\dots,a_k) \:\mid\: \{a_1,\dots,a_k\}\subseteq {\cal H}_P\}\]

\begin{definition}
An \CASP\ rule is an expression of the form
\begin{equation} \label{agg-rule}
A \leftarrow C_1,\ldots,C_m,A_1,\ldots,A_n,\naf B_1, \dots, \naf B_k
\end{equation}
where $A,A_1, \dots, A_n, B_1, \dots, B_k$ are ASP-atoms, and
$C_1, \dots, C_m$ are aggregate atoms ($m\geq 0$, $n\geq 0$, $k\geq 0$).\footnote{For 
methods to handle negated aggregate atoms, the reader is referred to 
\cite{SonPT06}.}

An \CASP\ program is a collection of \CASP\ rules.
\end{definition}
For an \CASP\ rule $r$ of the form (\ref{agg-rule}), we use the following
notations:
\begin{list}{$\circ$}{\topsep=1pt \parsep=0pt \itemsep=1pt}
\item  $head(r)$ denotes the ASP-atom $A$,
\item  $agg(r)$ denotes the set $\{C_1,\dots,C_m\}$,
\item  $pos(r)$ denotes the set $\{A_1,\ldots,A_n\}$,
\item $neg(r)$ denotes the set $\{B_1,\ldots,B_k\}$,
\item $body(r)$ denotes the right hand side of the rule
$r$. 
\end{list}
For a program $P$, $lit(P)$ denotes the set of all ASP-atoms 
present in $P$.

The syntax has been defined in such a way that collected and 
local variables of an aggregate atom $\ell$ have a scope that 
is limited to $\ell$. Thus, given an \CASP\ rule, it is possible
to rename these variables apart, so that each aggregate atom
$C_i$ in the body of a rule makes use of different collected and
local variables. Observe also that the collected and the local
variables are the only occurrences of variables from ${\cal V}_l$,
and these variables will not appear in any of $head(r)$, 
$pos(r)$, and $neg(r)$.

\begin{definition}
Given a term (atom, literal, rule) $\beta$, we denote
with $fvars(\beta)$ the set of variables from $\cal V$
present in $\beta$. We will refer to these as the 
\emph{free variables} of $\beta$.
The entity $\beta$ is \emph{ground} if $fvars(\beta)=\emptyset$.
\end{definition}
In defining the semantics of the language, we will need to 
consider all possible ground instances of programs. A
\emph{ground substitution} $\theta$ is a set 
$\{X_1 / a_1, \dots, X_k / a_k \}$, where the $X_i$ are
distinct elements of $\cal V$ and the elements $a_j$ are
constants from ${\cal F}_P$. Given a substitution $\theta$
and an ASP-atom (or an aggregate atom) $p$, the notation
$p\theta$ describes the atom obtained by simultaneously
replacing each occurrence of $X_i$ ($1\leq i \leq k$) with
$a_i$. The resulting element $p\theta$ is the \emph{instance}
of $p$ w.r.t. $\theta$. 

Given a rule $r$ of the form (\ref{agg-rule}) with
$fvars(r) = \{X_1, \dots, X_n\}$, and given a ground
substitution $\theta = \{X_1 / a_1, \dots, X_n / a_n\}$, the
ground instance of $r$ w.r.t. $\theta$ is the rule obtained from
$r$ by simultaneously replacing every occurrence of $X_i$
($i = 1,\dots,n$) in $r$ with $a_i$.

We will denote with $ground(r)$ the set of all the possible
ground instances of a rule $r$ that can be constructed 
in $\Sigma_P$.  For a program $P$, we will denote with $ground(P)$
the set of all ground instances of all rules in $P$, i.e., 
$ground(P) = \bigcup_{r\in P} ground(r)$. 

Observe that a ground 
logic program with aggregates differs from a ground logic 
program, in that it might still contain some local variables, 
which are members of ${\cal V}_l$, and they occur only in aggregate atoms. 

\begin{example}
Let ${\cal V} = \{Y\}$, ${\cal V}_l = \{X\}$, ${\cal F}_P=\{1,2,-2\}$, 
and $\Pi_P = \{p, q\}$. 
Let $r$ be the rule 
$$
\begin{array}{lll}
q(Y) & \leftarrow & \textnormal{\sc Sum}(\{X \mid p(X,Y)\}) \ge 0.
\end{array}
$$
$ground(r)$ will contain the following rules:
\[
\begin{array}{lll}
q(1) & \leftarrow & \textnormal{\sc Sum}(\{X \mid p(X,1)\}) \ge 0. \\
q(2) & \leftarrow & \textnormal{\sc Sum}(\{X \mid p(X,2)\}) \ge 0. \\
q(-2) & \leftarrow & \textnormal{\sc Sum}(\{X \mid p(X,-2)\}) \ge 0. \\
\end{array}
\]
Furthermore, for the aggregate atom $\ell = \textnormal{\sc Sum}(\{X \mid p(X,1)\}) \ge 0$, 
we have that 
\[
{\cal H}(\ell) = \{p(1,1), p(2,1), p(-2,1)\}.
\]
\hfill{$\Box$}
\end{example}

\section{Aggregate Solutions and Unfolding Semantics}
\label{semantics}
In this section, we develop our first characterization of the
semantics of program with aggregates, based on answer sets,
study some of its properties,
and investigate an implementation based on the {\sc Smodels} system.

\subsection{Solutions of Aggregates}
Let us start by developing the notion of interpretation, 
following the traditional structure~\cite{lloyd}.

\begin{definition}[Interpretation Domain]
The \emph{domain} $\cal D$ of an interpretation is the set
${\cal D} = {\cal H}_P \cup 2^{{\cal H}_P} \cup {\cal M}({\cal H}_P)$,
where $2^{{\cal H}_P}$ is the set of all (finite) subsets of
${\cal H}_P$, while ${\cal M}({\cal H}_P)$ denotes the set of
all finite multisets built using elements from ${\cal H}_P$.
\end{definition}

\begin{definition}[Interpretation]
An \emph{interpretation} $I$ is a pair $\langle {\cal D}, (\cdot)^I\rangle$,
where $(\cdot)^I$ is a function that maps ground terms to 
elements of $\cal D$ and ground atoms to truth values. The interpretation
function $(\cdot)^I$ is defined as follows:
\begin{itemize}
\item if $c$ is a constant, then $c^I = c$
\item if $s$ is a ground intensional set $\{X\:\mid\: q\}$, then
	$s^I$ is the set $\{a_1,\dots,a_k\} \in 2^{{\cal H}_P}$, where 
	$(\:q \{X / b\} \:)^I$ is true if and only if $b \in \{a_1,\dots,a_k\}$.

\item  if $s$ is a ground intensional multiset 
	$\{\!\!\{X\:\mid\: \exists \bar{Z} . q\}\!\!\}$, then
	$s^I$ is the multiset $\{\!\!\{a_1,\dots,a_k\}\!\!\} \in {\cal M}( {\cal H}_P )$, where,
	for each $i=1,\dots k$, there exists a ground substitution $\eta_i$ for $\bar{Z}$
	such that $(\: q (\eta_i \cup \{X / a_i\}) \:)^I$ is true,  and no other element
	has such property.

\item given an aggregate term $f(s)$,  then $f(s)^I$ is equal to $f^I(s^I)$, where 
\[ f^I \: : \: 2^{ {\cal H}_P } \cup {\cal M}( {\cal H}_P ) \: \rightarrow \: {\cal F}_P \]
	
\item if $p(a_1,\dots,a_k)$ is a ground ASP-atom or a ground aggregate atom, then $p(a_1,\dots, a_k)^I$ is 
	$p^I(a_1^I, \dots, a_k^I)$, where $p^I: {\cal D}^k \:\rightarrow\: \{true,false\}$.
\end{itemize}
\end{definition}
In the characterization of the aggregate functions, in this work we will mostly focus
on functions that maps sets/multisets to integer numbers in $\mathbb{Z}$. We will also
assume that the traditional aggregate functions and predicates are interpreted 
in the usual manner. E.g., $\textnormal{\sc SUM}^I$ is the function that sums the elements of
a set/multiset, and $\leq_7^I$ is the predicate that is true if its argument is an
element of $\mathbb{Z}$ no greater  than $7$.

Given a literal $not\:p$, its interpretation $(not\:p)^I$ is true (false) iff
$p^I$ is false (true).

For the sake of simplicity, given an atom (literal, aggregate atom) $p$, we will
denote with $I \models p$ the fact that  $p^I$ is true.

\begin{definition}[Rule Satisfaction]
Let $I$ be an interpretation and $r$ an \CASP\ rule. $I$ satisfies
the body of the rule ($I \models body(r)$) if $I\models q$ for each
$q \in body(r)$.
We say that $I$ satisfies $r$ if $I \models head(r)$ whenever
$I \models body(r)$.
\end{definition}
Finally, we can define the concept of model of a program.
\begin{definition}[Model of a Program]
An interpretation $I$ is a \emph{model} of a program $P$ if
$M$ satisfies each rule $r\in ground(P)$.
\end{definition}
In the rest of this work, we will assume  that the interpretation
of the aggregate functions and predicates is \emph{fixed}---i.e., it
is the same in all the  interpretations. This
allows us to keep the
``traditional'' view of  interpretations as subsets of ${\cal B}_P$
\cite{lloyd}.

\begin{definition}
[Minimal Model]
An interpretation 
$I$ is a \emph{minimal model} of $P$ if $I$ is a model of $P$ and 
there is no proper subset of $I$ which is also a model of $P$.
\end{definition}
We will now present the notion of \emph{solution of an aggregate}.
One of the guiding principles behind this concept is the following
observation. The satisfaction of an ASP-atom $p$ is \emph{monotonic}, in 
the sense that if $I\models p$ and $I \subseteq I'$, then we have
that $I'\models p$. This property does not hold any longer when
we consider aggregate atoms. Furthermore, the truth value of an 
aggregate atom $\ell$ depends on the truth value of certain atoms 
belonging to ${\cal H}(\ell)$. For example, if we consider the
aggregate atom $\ell = \textsc{SUM}(\{X\:\mid\:p(X)\}) \leq 1$ in 
the program with ${\cal H}(\ell) = \{p(1),p(2),p(-1)\}$, 
we can observe
that 
\[\begin{array}{lcl}
\{p(1)\} & \models & \textsc{SUM}(\{X\:\mid\:p(X)\}) \leq 1\\
\{p(1),p(2)\} & \not\models & \textsc{SUM}(\{X\:\mid\:p(X)\}) \leq 1
  \end{array}
\]
and $\ell$ is true if $p(2)$ is false or $p(-1)$ is true. These two 
observations lead to the following definition. 

\begin{definition}[Aggregate Solution]
Let $\ell$ be a ground aggregate atom. A {\em solution} of $\ell$
is a pair $\langle S_1, S_2 \rangle$ of  
disjoint subsets of ${\cal H}(\ell)$
such that, 
for every interpretation $I$, 
if $S_1 \subseteq I$ and 
$S_2 \cap I  = \emptyset$ then 
$I \models \ell$. 

We will denote with
${\cal SOLN}(\ell)$ the set of all the solutions
of the aggregate atom $\ell$.
\end{definition}
Let $S = \langle S_1, S_2 \rangle$ be the solution of an aggregate 
$\ell$; we denote with $S.p$ and $S.n$ the two
components 
$S_1$ and $S_2$ of the solution.
\begin{example} \label{ex-sol}
Let $c$ be the aggregate atom
${\textnormal{\sc Sum}(\{X \mid p(X)\}) {\ne} 5}$ 
in a language where ${\cal H}(c) =  \{p(1),p(2),p(3)\}$.
This aggregate atom has a total of $19$ solutions of the form 
$\langle S_1, S_2 \rangle$ such that 
$S_1, S_2 \subseteq \{p(1),p(2),p(3)\}$, 
$S_1 \cap S_2 = \emptyset$, and 
(i) either $p(1) \in S_1$; or (ii) $\{p(2), p(3)\} \cap  S_2 \ne \emptyset$.
These solutions are listed below.

\[\begin{array}{lclcl}
\langle \{p(1)\}, \emptyset\rangle & \hspace{.3cm} & \langle \{p(1)\}, \{p(2)\}\rangle & \hspace{.3cm} & \langle \{p(1)\}, \{p(3)\}\rangle\\ 
\langle \{p(1)\}, \{p(2),p(3)\}\rangle & &\langle \{p(1), p(2)\}, \emptyset\rangle && \langle \{p(1), p(2)\}, \{p(3)\}\rangle \\
\langle \{p(1), p(3)\}, \emptyset\rangle && \langle \{p(1), p(3)\}, \{p(2)\}\rangle && \langle \{p(2)\}, \{p(3)\}\rangle \\
\langle \{p(2)\}, \{p(3),p(1)\}\rangle && \langle \{p(3)\}, \{p(2)\}\rangle && \langle \{p(3)\}, \{p(2),p(1)\}\rangle \\
\langle \{p(1), p(2), p(3)\}, \emptyset\rangle 
	&& \langle \emptyset, \{p(2)\} \rangle 
	&& \langle \emptyset, \{p(3)\} \rangle\\
\langle \emptyset, \{p(1),p(2)\}\rangle && 
	\langle \emptyset, \{p(1),p(3)\}\rangle &&
	\langle \emptyset, \{p(2),p(3)\}\rangle\\
\langle \emptyset, \{p(1),p(2),p(3)\}\rangle && && 
  \end{array}
\]
\hfill $\Box$
\end{example}
Let $\ell$ be an aggregate atom. 
The following properties hold:
\begin{observation} \label{obs0}
\hspace{1cm}
\begin{list}{}{\topsep=1pt \parsep=0pt \itemsep=1pt \leftmargin=10pt}
\item[{\em (i)}] If there is at least one interpretation $I$ such
that $I \models \ell$, then
${\cal SOLN}(\ell) \ne \emptyset$.

\item[{\em (ii)}] If $S_{\ell}$ is a solution of $\ell$ then, 
for every set $S' \subseteq {\cal H}(\ell)$ with 
$S' \cap (S_\ell.p \cup S_\ell.n) = \emptyset$, we have that
$\langle S_\ell.p, S_\ell.n \cup S'\rangle$ and 
$\langle S_\ell.p \cup S', S_\ell.n \rangle$
are also solutions of $\ell$.
\end{list}
\end{observation}
The first property holds since the pair
$\langle I \cap {\cal H}(\ell), {\cal H}(\ell) \setminus I\rangle$
is a solution of $\ell$. The second property is trivial from the 
definition of a solution.

\subsection{\CASP\ Answer Sets}
We will now define the
\emph{unfolding} of an aggregate atom,
of a ground rule, and of a program. For simplicity,
we use $S$ (resp. $\naf S$) to denote the conjunction 
$\bigwedge_{a \in S} a$ (resp. $\bigwedge_{b \in S} \naf b$)
when $S \ne \emptyset$; 
$\emptyset$ ($\naf \emptyset$) stands for 
$\top$ ($\bot$).\footnote{We follow the convention of
denoting $true$ with $\top$ and $false$ with $\bot$.} 
\begin{definition}[Unfolding of an Aggregate Atom]
Given a ground aggregate atom $\ell$
and a solution $S \in {\cal SOLN}(\ell)$, 
the unfolding of $\ell$ w.r.t. $S$, denoted by $\ell(S)$, is 
$S.p \; \wedge\; \naf S.n$.  
\end{definition}
\begin{definition}[Unfolding of a Rule] \label{unfolding}
Let $r$ be a ground rule 
\[
A \leftarrow C_1,\ldots,C_m,A_1,\ldots,A_n,\naf B_1, \dots, \naf B_k
\]
where $\langle C_i \rangle_{i=1}^m$ are aggregate atoms. 
A ground rule $r'$ is an {\em unfolding} of $r$ if 
there exists a sequence of aggregate solutions $S_{C_1}, \dots,
S_{C_m}$  such that
\begin{enumerate}
\item $S_{C_i}$ is a solution of the aggregate atoms $C_i$ ($i=1,\dots,m$),
\item $head(r') = head(r)$, 
\item $pos(r') = pos(r) \cup \bigcup_{i=1}^m S_{C_i}.p$,  
\item $neg(r') = neg(r) \cup \bigcup_{i=1}^m S_{C_i}.n$, and 
\item $agg(r') = \emptyset$. 
\end{enumerate}
We say that $r'$ is \emph{an unfolding of $r$ 
with respect to} $\langle S_{C_i} \rangle_{i=1}^m$. The set
of all possible unfoldings of a rule $r$ is denoted by
$unfolding(r)$.
\end{definition} 
For an \CASP\ program $P$, $unfolding(P)$ denotes the set of 
the unfoldings of the rules in $ground(P)$. It is easy to see that 
$unfolding(P)$ is a normal logic program.

The answer sets of \CASP\ programs  are defined as follows.
\begin{definition}
\label{answerset}
A set of atoms $M$ is an \CASP-answer set of $P$ iff 
$M$ is an answer set of $unfolding(P)$.
\end{definition}
\begin{example} \label{exp1}
Let $P_1$ be the program:\footnote{
   We would like to thank Vladimir Lifschitz for suggesting this example. 
}
\[
\begin{array}{lcl}
                  p(a) & \leftarrow & \textnormal{\sc Count}(\{X \mid p(X)\}) > 0 \\
                  p(b) & \leftarrow & \naf q \\
                  q    & \leftarrow & \naf p(b) \\
\end{array}
\]
The aggregate atom
$\textnormal{\sc Count}(\{X \mid p(X)\}) > 0$ has five aggregate solutions:
\[\begin{array}{lclclclcl}
\langle \{p(a)\}, \emptyset \rangle & \hspace{.25cm} & 
\langle \{p(b)\}, \emptyset \rangle & \hspace{.25cm} & 
\langle \{p(a),p(b)\}, \emptyset \rangle &
\langle \{p(a)\}, \{p(b)\} \rangle & \hspace{.25cm} & 
\langle \{p(b)\}, \{p(a)\} \rangle & \hspace{.25cm} 
  \end{array}
\]
The unfolding of $P_1$ is the program 
\[ \begin{array}{lclclcl}
  p(a)& \leftarrow &  p(a)  & \hspace{1cm}& 
  p(a)& \leftarrow &  p(b)  \\
  p(a)& \leftarrow &  p(a),p(b) &     &             
  p(a)& \leftarrow &  p(a),\naf p(b) \\
  p(b)& \leftarrow &  \naf q &  & 
     q& \leftarrow &  \naf p(b) \\
  p(a)& \leftarrow &  p(b),\naf p(a) 	
\end{array}
\]
$M_1 = \{q\}$ and 
$M_2 = \{p(b), p(a)\}$ are the two answer sets of $unfolding(P_1)$,
thus \CASP-answer sets of $P_1$. \hfill $\Box$
\end{example}
\begin{example} \label{exp2}
Let $P_2$ be the program 
\[
\begin{array}{lclclcl}
   p(1) & \leftarrow \\
   p(2) & \leftarrow \\
   p(3) & \leftarrow\\
   p(5) & \leftarrow & q\\
   q & \leftarrow& \textnormal{\sc Sum}(\{X \mid p(X)\}) > 10 
\end{array}
\]
The only aggregate solution of $\textnormal{\sc Sum}(\{X \mid p(X)\}) > 10$ 
is 
$\langle \{p(1), p(2), p(3), p(5)\} , \emptyset \rangle$ 
and $unfolding(P_2)$ contains:
\[
\begin{array}{lclcl}
 p(1) & \leftarrow \\
 p(2) & \leftarrow \\
 p(3)&\leftarrow \\
 p(5) & \leftarrow &  q\\
 q & \leftarrow &  p(1),p(2),p(3),p(5) 
\end{array}
\]
which has $M_1 = \{p(1),p(2),p(3)\}$ as its only answer set. 
Thus, $M_1$ is the only \CASP-answer set of $P_2$. \hfill $\Box$
\end{example} %
The next program with aggregates  does not 
have answer sets, even though it does not contain any
negation as failure literals.
\begin{example}
Consider the program $P_3$:
\[
\begin{array}{lcl}
p(2) &\leftarrow &\\
p(1) & \leftarrow &  \textnormal{\sc Min}(\{X \mid p(X)\}) \ge 2
\end{array}
\]
The unique aggregate solution of the aggregate atom
$\textnormal{\sc Min}(\{X \mid p(X)\}) \ge 2 $ 
with respect to ${\cal B}_{P_3} = \{p(1),p(2)\}$
is 
$\langle \{p(2)\}, \{p(1)\} \rangle$.
The unfolding of $P_3$ consists of the two rules:
\[
\begin{array}{lcl}
p(2) & \leftarrow &\\
p(1) & \leftarrow  & p(2), \naf p(1)
\end{array}
\]
and it  does not have any answer sets. 
As such, $P_3$ does not have any \CASP-answer sets. \hfill $\Box$
\end{example}
%
Observe that, in creating $unfolding(P)$, we use 
\emph{every} solution of $c$ in ${\cal SOLN}(c)$. Since the number 
of solutions of an aggregate atom can be exponential in
the size of the Herbrand base, the size of $unfolding(P)$ 
can be exponential in the size of $P$. Fortunately, as we 
will show later (Theorem \ref{theorem10}), this process can be 
simplified by considering only minimal solutions of each 
aggregate atom
(Definition \ref{def9}). 
In practice, for most common uses of aggregates, we have observed
a small number of elements in the minimal solution set (typically
linear or quadratic  in the extension of the predicate used in the intensional
set).


\subsection{Properties of \CASP-Answer Sets}
\label{sec-prop}

It is easy to see that the notion of \CASP-answer sets 
extends the notion of answer sets of normal logic programs.
Indeed, if $P$ does not contain 
aggregate atoms, then $unfolding(P)=ground(P)$. Thus, 
for a program without aggregates $P$, 
$M$ is an \CASP-answer set of $P$ if and only if $M$ 
is an answer set of $P$ with respect to the Gelfond-Lifschitz 
definition of answer sets. 

We will now show that \CASP-answer sets satisfies the same properties of
 minimality, closedness, and supportedness as answer sets for normal
logic programs. 

\begin{lemma}
\label{lem1}
Every model of $unfolding(P)$ is a model 
of $P$.
\end{lemma}
\noindent {\bf Proof.}
Let $M$ be a model of $unfolding(P)$, and let us 
consider a rule $r \in ground(P)$ such that $M$ satisfies the 
body of $r$. This implies that 
there exists a sequence of solutions 
$\langle S_c \rangle_{c \in agg(r)}$
for the
aggregate atoms occurring in $r$,
such that 
$S_c \in {\cal SOLN}(c)$, $S_c.p \subseteq M$, 
and $S_c.n \cap M = \emptyset$. Let $r'$ be 
the unfolding of $r$ with respect to $\langle S_c \rangle_{c \in agg(r)}$.
We have that $pos(r') \subseteq M$ and $neg(r') \cap M = \emptyset$.
In other words, $M$ satisfies the body of $r' \in unfolding(P)$.
This implies that $head(r') \in M$, i.e., $head(r) \in M$.
\hfill$\Box$

\begin{lemma}
\label{lem2}
Every model of $P$ is a model of $unfolding(P)$.
\end{lemma}
\noindent {\bf Proof.}
Let $M$ be a model of $P$, and let us 
consider a rule $r' \in unfolding(P)$ such that $M$ satisfies 
the body of $r'$. Since $r' \in unfolding(P)$, there exists $r \in ground(P)$
and a sequence of aggregate solutions $\langle S_c \rangle_{c \in agg(r)}$
for the aggregate atoms in $r$ such that $M$ satisfies $S_c.p \wedge \naf S_c.n$ 
(for $c \in agg(r)$)
and $r'$ is the unfolding of $r$ with respect to $\langle S_c \rangle_{c \in agg(r)}$. 
This means that $pos(r) \subseteq M$, $neg(r) \cap M = \emptyset$,
and $M \models c$ for $c \in agg(r)$. In other words, $M$ satisfies
$body(r)$. Since $M$ is a model of $ground(P)$, we have that 
$head(r) \in M$, which means that $head(r') \in M$. 
\hfill$\Box$

\begin{theorem}
\label{th1}
Let $P$ be a program with aggregates and $M$ be an
\CASP-answer set of $P$. Then, $M$ is closed, supported,
and a minimal model of $ground(P)$.
\end{theorem} 
{\bf Proof.} Since $M$ is an \CASP-answer set 
of $P$, Lemma \ref{lem1} implies that $M$ is a model of $ground(P)$. 
Minimality of $M$ follows from Lemma \ref{lem2} 
and from the fact that $M$ is a minimal model of $unfolding(P)$. 
Closedness is immediate from Lemma~\ref{lem1}.

Supportedness can be derived from the fact that each
atom $p$ in $M$ is supported by
$M$ (w.r.t. $unfolding(P)$) since $M$ is an answer set of 
$unfolding(P)$. Thus, if $p$ were not supported by $M$ w.r.t.
$ground(P)$, then this would mean that no rule in 
$unfolding(P)$ supports $p$, which would contradict
the fact that $M$ is an answer set of $unfolding(P)$.
\qed

\medskip\noindent
Observe that the converse of the above theorem does not hold, as
illustrated by the following example.

\begin{example} \label{p4-ex}
Let $P_4$ be the program 
\[
\begin{array}{lcl}
 p(1) & \leftarrow & \\
  p(2) & \leftarrow &  q \\    
 q & \leftarrow &  \textnormal{\sc Sum}(\{X \mid p(X)\}) \ge 2 \\
   q & \leftarrow &  \textnormal{\sc Sum}(\{X \mid p(X)\}) < 2 
\end{array}
\]
It is easy to see that $M = \{p(1), p(2), q\}$ is a minimal 
model of this ground program---i.e., $M$ is
a minimal set of atoms, closed under the rules of $ground(P_4)$ 
and each atom of $M$ is supported by a rule of $ground(P_4)$. 
On the other hand, $unfolding(P_4)$ consists of the following 
rules
\[
\begin{array}{lllllllll}
 p(1) & \leftarrow  \\
 p(2) & \leftarrow &  q & \hspace{0.5cm} & 
 q  & \leftarrow &  p(1), p(2)\\
 q  & \leftarrow &  p(2) & &
 q  & \leftarrow &  p(2), \naf p(1)\\
  q & \leftarrow &  p(1), \naf p(2) & &   
  q & \leftarrow &  \naf p(1), \naf p(2) & \hspace{0.5cm} \\
q & \leftarrow & \naf p(2) \\
\end{array}
\]
$M$ is not an answer set of $unfolding(P_4)$. We can easily check that this 
program does not have an answer set. Thus, $P_4$ does not have an answer set  
according to Definition \ref{answerset}. \hfill $\Box$
\end{example}


\begin{remark}
{\rm 
The above result might seem counterintuitive, and it deserves some discussion. 
One might argue that, in any interpretation of the program $P_4$, either 
$$ \textnormal{\sc Sum}(\{X \mid p(X)\}) \ge 2 \:\:\:\:\: or 
\:\:\:\:\:
\textnormal{\sc Sum}(\{X \mid p(X)\}) < 2$$ will be true. 
As such, $q$ would appear to be  true, and hence $M$ should be an answer set 
of the program. 

While this is  a possible way to deal with 
aggregates, in this example, this line of reasoning might lead to
circular justifications of atoms in $M$. In fact,
observe that the rules that support $p(2)$ and $q$ in $M$ are 
$p(2) \leftarrow q$ and 
$q \leftarrow \textnormal{\sc Sum}(\{X \mid p(X)\}) {\ge} 2$,
respectively. In the context of the program, 
$\textnormal{\sc Sum}(\{X \mid p(X)\}) \ge 2$ can be true 
only if $p(2)$ is true. This is equivalent to say that 
$p(2)$ is true because  $q$ is true, and $q$ is true 
because  $p(2)$ is true. In other words, 
the answer set contains two elements whose
truth values depend on each other. 

The traditional answer set definition in \cite{GL88} does not allow such 
type of justifications---in that it does not consider $\{a\}$ as 
an answer set of the program $\{a \leftarrow a\}$.

Example \ref{p4-ex} shows that our 
approach to defining the semantics of logic programs with aggregates 
is closer to the spirit of the traditional 
answer set definition. 

We should also  observe that most of the   recent approaches to 
handling aggregates (e.g., \cite{faberLP04,ferraris,Pelov04})
yield the same result on this example. Moreover, if we encode $P_4$ 
in {\sc Smodels} (using weight constraints) as 
\begin{verbatim}
       p(1).    p(2).    q:- 2[p(1)=1, p(2)=2].    q:-[p(1)=1, p(2)=2]1.
\end{verbatim}
we obtain an {\sc Smodels} program that does not have any answer sets.
}
\end{remark}

\subsection{Implementation} 
\label{impl}

In spite of the number of proposals dealing with aggregates in logic 
programming, only few implementations have been described. Dell'Armi 
et al. \cite{armi-ijcai03} describe an implementation of aggregates 
in the {\sc dlv} engine, based on  the semantics described in 
Section \ref{old-semantics} (the current distribution is limited to 
aggregate-stratified programs\footnote{The concept of aggregate
stratification is discussed in Subsection~\ref{stra}.}). Elkabani et al. \cite{ElkabaniPS04} describe 
an integration of a Constraint Logic Programming  engine (the
ECLiPSe engine)
and the {\sc Smodels} answer set solver; the integration is employed to implement 
aggregates, with respect to the semantics of 
Section \ref{old-semantics}. Some more restricted forms of aggregation, 
characterized according to the semantics
of Section \ref{old-semantics} have also been introduced in 
the {\sc ASET-Prolog} system \cite{a-prolog}.
Efficient algorithms for bottom-up computation of
the perfect model of aggregate-stratified programs
have been described in \cite{KempR98,ZanioloAO93}.

In this section, we will describe an implementation of a system 
for computing \CASP-answer sets based on the computation of the  
solutions of aggregate atoms, unfolding of the program, and 
computation of the answer sets using an off-the-shelf answer set solver. 
We begin with a discussion of computing solutions of aggregate atoms. 

\subsubsection{Computing the Solutions}
As we have mentioned
before, the size of the program $unfolding(P)$
can become unmanageable in some situations. One way to 
reduce the size of $unfolding(P)$ is to find a set of ``representative'' 
solutions for the aggregate atoms occurring in $P$, whose size 
is---hopefully---smaller than the size of the ${\cal SOLN}(\ell)$. 
Interestingly, in several situations, the number of representative 
solutions of an aggregate atom is small \cite{SonP05}. We say that a 
set of solutions is complete if it can be used to check the 
satisfiability of the aggregate atom in every interpretation of 
the program. First, we define when a solution \emph{covers} another solution.

\begin{definition}
A solution $S$ of an aggregate atom $\ell$ \emph{covers} a solution
$T$ of $\ell$, denoted by $T \unlhd_{\ell} S$,
if, for all interpretations $I$,
$$ (\: I \models (T.p \wedge not\:T.n) \:) \:\:\Rightarrow \:\: (\: I 
\models (S.p \wedge not\:S.n)\:)$$
\end{definition}
This can be used to define a {\em complete} and {\em minimal} sets of solutions of an 
aggregate atom.

\begin{definition} \label{def9} A set $S(\ell)$ of solutions of 
an aggregate atom $\ell$ is 
{\em complete} if for every solution $S_\ell$ of $\ell$, there exists 
$T_{\ell} \in S(\ell)$ such that $S_\ell \unlhd_\ell  T_{\ell}$.

A solution set $S(\ell)$ is \emph{reducible} if there are two distinct 
solutions
$S$ and $T$ in $S(\ell)$ such that
$T \unlhd_\ell S$. The set of solutions
$S(\ell) \setminus \{T\}$ is then called  a \emph{reduction} of $S(\ell)$.
A solution set $S(\ell)$ is \emph{minimal} if it is complete and not reducible. 
\end{definition}

By definition, we have that ${\cal SOLN}(\ell)$ is complete. Because
of the transitivity of the covering relationship, we can conclude 
that any minimal solution set of $\ell$ is a reduction of ${\cal SOLN}(\ell)$. 
Given a ground  program $P$, let $c_1,\dots,c_k$ be the aggregate 
atoms present in $P$, and let us denote with 
$\zeta(P,[ c_1/S(c_1), \dots, c_k/S(c_k) ])$ the unfolding of $P$ 
where $c_i$ has been unfolded using only the solution set $S(c_i)$. 

\begin{theorem} \label{theorem10} 
Given a ground program $P$ containing the aggregate atoms
$c_1,\dots,c_k$, and given
a complete solution set $S(c_i)$ for each aggregate atom $c_i$, we have that $M$ is an 
\CASP-answer set of $P$ 
iff $M$ is an answer set of $\zeta(P,[c_1/S(c_1),\dots,c_k/S(c_k)])$.
\end{theorem}
\noindent {\bf Proof.}
For an interpretation $M$, let
$Q_1 = (\zeta(P,[c_1/S(c_1),\dots,c_k/S(c_k)]))^M$
and $Q_2 = (\zeta(P,[c_1/{\cal SOLN}(c_1),\dots,c_k/{\cal SOLN}(c_k)]))^M = (unfolding(P))^M$. 
We have that $M$ is an \CASP-answer set of $P$ iff $M$ is an answer set of $Q_2$. 
Furthermore, $Q_1 \subseteq Q_2$, and for each rule $r \in Q_2$ there is
a rule $r' \in Q_1$ with $head(r)=head(r')$ and $body(r') \subseteq body(r)$. 
Using this information, we can show that $M$ is an answer set of $Q_1$ iff $M$ 
is an answer set of $Q_2$, which proves the theorem. \hfill$\Box$

\noindent
The above theorem shows that
we can use any complete solution set (e.g., a minimal
one) to unfold an aggregate atom.

We make use of the following observation to compute
a complete solution set:
\begin{observation}
Let $\ell$ be an aggregate atom  and let
  $\langle S_1, S_2\rangle$, $\langle T_1,T_2\rangle$
be solutions of $\ell$. Then
$\langle T_1,T_2 \rangle \unlhd_\ell \langle S_1,S_2\rangle \mbox{  iff  }
	S_1 \subseteq T_1 \mbox{ and } S_2 \subseteq T_2$. 
\end{observation} 
The abstract algorithm in Figure \ref{al1} computes a complete 
solution set ${\cal S}(\ell)$ for a given aggregate atom---when 
called with {\tt Find\_Solution}$(\ell,\langle\emptyset,\emptyset\rangle)$ and
with initially ${\cal S}(\ell) =\emptyset$\/. This algorithm is generic---i.e., can be 
used with arbitrary aggregate predicates, as long as a mechanism to 
perform the test in line {\tt 3} is provided. The test is used to 
check whether the current $\langle T,F \rangle$ represents a solution of $\ell$.
Observe also that more effective algorithms can be provided for specific 
classes of aggregates, by using properties of the aggregate predicates 
used in the aggregate atoms \cite{SonP05}.

\begin{figure}[htb]
\begin{center}
\begin{minipage}[t]{.8\textwidth}
{\small\tt \begin{tabbing} 
iiii\=iiii\=iiiii\=iiii\=iiii\=iiii\=iiii\=iiii\=iiii\kill
1:\>{\bf Procedure} {\tt Find\_Solution} ($\ell$, $\langle T, F\rangle$)\\ 
2:\>\{ assume $T = \{t_1,\dots,t_k\}$ and $F=\{f_1,\dots,f_h\}$ \}\\ 
3:\>\> {\bf if} $t_1\wedge \dots \wedge t_k \wedge \neg f_1 \wedge \dots \wedge \neg f_h \models \ell$ {\bf then}\\ 
4:\>\>\>  Add $\langle T,F \rangle$ to ${\cal S}(\ell)$; \\ 
5:\>\>\> {\bf return}\\ 
6:\>\> {\bf endif}\\ 
7:\>\> {\bf if} $T \cup F = {\cal B}_P$ {\bf then return};\\ 
8:\>\> {\bf endif}\\ 
9:\>\> {\bf forall} ($p \in {\cal B}_P \setminus (T \cup F)$)\\ 
10:\>\>\> {\tt Find\_Solution}($\ell$, $\langle T\cup \{p\}, F\rangle$);\\ 
11:\>\>\> {\tt Find\_Solution}($\ell$, $\langle T, F \cup \{p\}\rangle$); \\
12:\>\> {\bf endfor} \end{tabbing}} \end{minipage}\end{center} 
\caption{Algorithm to compute solution set of an aggregate} \label{al1} \end{figure}

Given a program $P$ containing the aggregate atoms $c_1,\dots,c_k$, we can replace $P$ with $P'=\zeta(P, [c_1/{\cal 
S}(c_1), \dots
	c_k/{\cal S}(c_k)])$. The  program $P'$ is a
normal logic program without aggregates, whose answer sets can be computed using a standard answer set solver. 
The algorithm has been implemented
in an extended version of {\sc lparse}---using an external 
constraint solver to compute line {\tt 3}. Note that the {\bf forall} in line {\tt 9} 
is a \emph{non-deterministic} choice of $p$.

\subsubsection{The \CASP\ System}
\begin{figure}[htb]
\begin{center}
\centerline{\psfig{figure=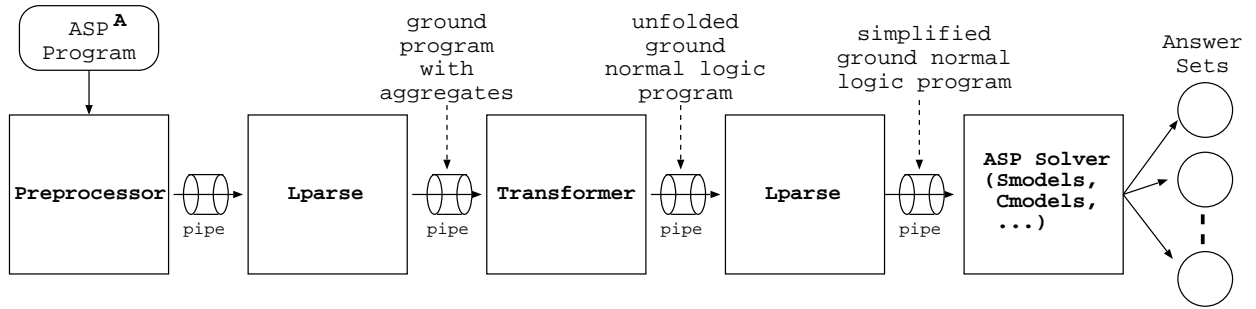,width=\textwidth}}
\caption{Overall System Structure}
\label{big}
\end{center}
\end{figure}

We will now describe the prototype we have
constructed, called \CASP,
for computing answer sets of
programs with aggregates. The computation is performed following
the semantics given in Definition \ref{answerset}, 
simplified by Theorem \ref{theorem10}. 
In other words, to compute the answer set of a program $P$, we 
\begin{enumerate} 
\item Compute a complete (and possibly minimal) solution set for each 
aggregate atom occurring in  $P$; 

\item  Unfold $P$ using the computed solution sets;

\item Compute the answer sets of the unfolded program $unfolding(P)$ using a standard answer
	set solver (in our case, both {\sc Smodels} and {\sc Cmodels}).
\end{enumerate}
The overall structure of the
system is shown in Figure ~\ref{big}.

The computation of answer sets is performed in five steps. In the first 
step, a preprocessor performs a number of simple syntactic 
transformations on the input program, which are aimed at 
rewriting the aggregate atoms in a format acceptable by {\sc lparse}. For example, 
the aggregate atom $ \textnormal{\sc Sum}(\{X \:\mid\: p(X)\}) \geq 40 $ is rewritten to 
$ \textnormal{``}\$agg\textnormal{''} (sum, 
\textnormal{``}\$x\textnormal{''}, p(\textnormal{``}\$x\textnormal{''}), 
40, geq)
$
and an additional rule 
\[
\textnormal{\tt 0\:\{``\$agg''(sum,``\$x'', p(``\$x''), 40, geq)\}\:1}
\] 
is added to the program.
The rewritten program is then grounded and simplified using {\sc lparse}, 
in which aggregate atoms are treated like standard (non-aggregate) literals.

\begin{figure}[htb] \centerline{\psfig{figure=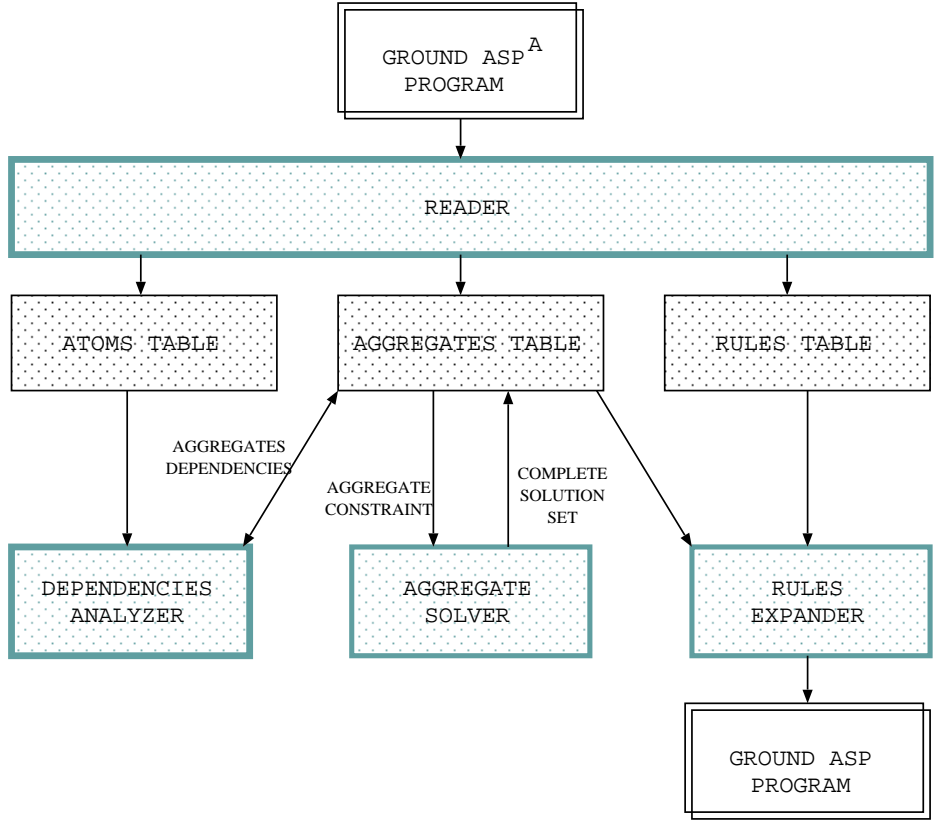,width=.75\textwidth}}
\caption{Transformer Module}
\label{transformer}
\end{figure}

The ground program is processed by the {\em transformer module}, 
detailed in Figure ~\ref{transformer}, in which the unfolded program is 
computed. This module performs the following
operations:
\begin{enumerate}
\item Creation of  the \emph{atom table}, the
	\emph{aggregate table},
	and  the \emph{rule table}, used to store the ground atoms,
	aggregate atoms, and rules of the program, respectively. This
	is performed by the {\em Reader} component in Figure
	\ref{transformer}.
\item Identification of  the dependencies between
aggregate atoms and the atoms contributing to such atoms
(done by the {\em Dependencies Analyzer});
\item Computation of a complete solution set for
each aggregate atom (done by the {\em Aggregate Solver}---as
	described in the previous subsection);
\item Creation of the unfolded program
(done by the {\em Rule Expander}).
\end{enumerate}
Note that the unfolded program is passed one more time through {\sc lparse}, 
to avail of the simplifications and optimizations that {\sc lparse} can 
perform on a normal logic program (e.g., expansion of domain predicates
and removal of unnecessary rules). The resulting program is  a ground 
normal logic program, whose answer sets can be computed by a system 
like {\sc Smodels} or {\sc Cmodels}.

\subsubsection{Some Experimental Results}

We have performed a number of tests using the \CASP\ 
system. In particular, we selected  benchmarks 
with aggregates presented in the literature. 
The benchmarks, drawn from various papers on aggregation, are:
\begin{itemize}
\item \emph{Company Control:} Let $owns(X,Y,N)$ denotes the fact that company $X$ owns a fraction
$N$ of the shares of the company $Y$. We say that a company $X$ 
\emph{controls} a company $Y$ if the sum of the shares it owns in $Y$
together with the sum of the shares owned in $Y$ by companies 
controlled by $X$ is greater than half of the total shares of 
$Y$:
\[\begin{array}{lcl}
  control\_shares(X, Y, N) & \leftarrow & owns(X,Y,N) \\
  control\_shares(X, Y, N)& \leftarrow &control(X,Z), owns(Z,Y,N) \\
  control(X,Y)& \leftarrow & \textsc{Sum}(\{\!\!\{\: M \:|\: control\_shares(X,Y,M)\:\}\!\!\}) > 50\\
  \end{array}
\]
We explored different instances, with varying numbers of companies.

\item \emph{Shortest Path:}
Suppose a weight-graph is given by relation $arc$, where $arc(X,Y,W)$ means 
that there is an arc in the graph from node $X$ to node $Y$ of weight $W$. We represent 
the shortest path (minimal weight) relation $spath$ using the following rules
\[\begin{array}{lcl}
path(X,Y,C) & \leftarrow &  arc(X,Y,C) \\
path(X,Y,C) & \leftarrow &  spath(X,Z,C1), arc(Z,Y,C2), C = C1 + C2 \\
spath(X,Y,C) & \leftarrow & \textsc{Min}(\{\!\!\{\: D \:|\: path(X,Y,D)\:\}\!\!\}) = C \\
  \end{array}
\]
The instances explored make use of graphs with varying number of nodes.

\item \emph{Party Invitations:}
The main idea of this problem is to send out party invitations considering 
that some people will not accept the invitation unless they know that at 
least $k$ other people from their friends accept it too.
\[\begin{array}{lcl}
friend(X,Y)& \leftarrow & friend(Y,X)\\
coming(X)& \leftarrow & requires(X,0)\\
coming(X)& \leftarrow & requires(X,K), \textsc{Count}(\{\: Y \:|\: come\_friend(X,Y)\:\}) \geq K\\
come\_friend(X,Y) &   \leftarrow & friend(X,Y), coming(Y)\\
  \end{array}
\]
The instances explored in our experiments have different numbers
of people invited to the party.

\item \emph{Group Seating:}
In this problem, we want to arrange the sitting of a group of $n$ people  
in a restaurant, knowing that the number of tables times the number of 
seats on each table equals to $n$. 
The number of people that can sit at a table cannot exceed the number 
of chairs at this table, and each person can sit exactly at one table. 
In addition, people who like each other must 
sit together at the same table and those who dislike 
each other must sit at different tables.
\[\begin{array}{lcl}
at(P,T)& \leftarrow & person(P), table(T), not \:not\_at(P,T)\\
not\_at(P,T)& \leftarrow & person(P), table(T), not\: at(P,T)\\
 & \leftarrow & table(T), nchairs(C), \textsc{Count}(\{\:P \:|\: at(P,T)\:\}) > C\\
 & \leftarrow & person(P), \textsc{Count} ( \{\:T \:|\: at(P,T)\:\}) \neq 1\\
 & \leftarrow & like(P1,P2), at(P1,T), not\: at(P2,T)\\
 & \leftarrow & dislike(P1,P2), at(P1,T), at(P2,T)
\end{array}
\]
The benchmark makes use of 16 guests, 4 tables, each having 4 chairs.
 
\item \emph{Employee Raise:}
Assume that a manager decides to select at most $N$ employees to
give them a raise. An employee is a good candidate for the raise if he has worked 
for at least $K$ hours per week. 
A relation $emp(X,D,H)$ denotes that an employee $X$ worked $H$ hours during the day $D$.
\[\begin{array}{lcl}
raised(X)& \leftarrow & empName(X), not\: notraised(X) \\
notraised(X)& \leftarrow & empName(X), not\: raised(X) \\
notraised(X)& \leftarrow & empName(X), nHours(K), 
	\textsc{Sum}(\{\!\!\{H \:|\: emp(X,D,H)\:\}\!\!\}) < K\\
 & \leftarrow & maxRaised(N), \textsc{Count}(\{X \:|\:raised(X)\}) > N
  \end{array}
\]
The different experiments conducted are described by the two parameters
$M/N$, where $M$ is the number of employees and $N$ the maximum number
of individuals getting a raise.

\item \emph{NM1} and \emph{NM2}: these are two synthetic benchmarks that compute large 
	aggregates that are recursive and non-monotonic. \emph{NM1} has its core in the 
	following rules:
\[\begin{array}{lcl}
q(K) & \leftarrow & r(X), w(K), max({X \:|\: p(X)})=K\\
p(X) & \leftarrow & q(K), r(X), w(K)\\
a(X) & \leftarrow & not b(X), p(X),r(X)\\
b(X) & \leftarrow & not a(X), p(X),r(X)
  \end{array}
\]
The program \emph{NM2} relies on the following set of rules:
\[
\begin{array}{lcl}
q(K) & \leftarrow & r(X), w(K), min({X \:|\:p(X)})>K\\
p(X) & \leftarrow & q(K), r(X), w(K)
\end{array}
\]
\end{itemize}
The code for the benchmarks can be found at:
\url{www.cs.nmsu.edu/~ielkaban/asp-aggr.html}.

Table~\ref{tab1} presents the results obtained. 
The columns of the table have the following meaning: 
\begin{itemize} 
\item {\tt Program} is the name of the benchmark. 
\item {\tt Instance} describes the specific instance of the
	benchmark used in the test. 
\item {\tt Smodels Time} is the time (in seconds) employed by {\sc Smodels} to
	compute the answer sets of the unfolded program.
\item {\tt Cmodels Time} is the time (in seconds) employed by {\sc Cmodels} to
	compute the answer sets of the unfolded program.
\item {\tt Transformer Time} is the time (in seconds) to preprocess and ground the
	program (i.e., compute the solutions of aggregates and perform
	the unfolding---this includes the complete pipeline discussed in
	Figure~\ref{big}).
\item {\tt DLV$^{\mathcal{A}}$} is the time employed by the
	{\tt DLV$^{\mathcal{A}}$} system to execute the same benchmark (where
	applicable, otherwise marked {\tt N/A})---observe that the current distribution of this
	system does not support recursion through aggregates.
\end{itemize}
All computations have been performed on a Pentium 4, 3.06 GHz machine 
with 512MB of memory under Linux 2.4.28 using GCC 3.2.1. 
The system is available for download at
\url{www.cs.nmsu.edu/~ielkaban/asp-aggr.html}.

As we can see from the table, even this relatively simple implementation
of aggregates can efficiently solve all benchmarks we tried, offering a 
coverage significantly larger than other existing implementations. Observe also
that the overhead introduced by the computation of aggregate solutions is significant
in very few cases.

\begin{table}[htbp]
\begin{center}
{\footnotesize
\begin{tabular}{|l||c|c|c|c|c|}
\hline
Program            & Instance & Smodels  & Cmodels   & Transformer   & {\tt DLV$^{\mathcal{A}}$}    \\
                    &          & Time         & Time          & Time            & Time       \\
                    &          &          &           & {\small (Preprocessor, Lparse,}            &        \\
                    &          &          &           & {\small Transformer, and Lparse)}            &        \\
\hline
Company Control  & 20           &  0.010        &  0.00          & 0.080         & N/A     \\
Company Control  & 40           &  0.020        &  0.00          & 0.340         & N/A     \\
Company Control  & 80           &  0.030        &  0.00          & 2.850         & N/A     \\
Company Control  & 120          &  0.040        &  0.030         & 12.100        & N/A     \\
Shortest Path     & 20           &  0.220        &  0.05          & 0.740        & N/A      \\
Shortest Path     & 30           &  0.790        &  0.13          & 2.640       & N/A      \\
Shortest Path     & 50           &  3.510        &  0.51          & 13.400      & N/A    \\
Shortest Path (All Pairs)     & 20           &  6.020  & 1.15   & 35.400        & N/A    \\
Party Invitations  & 40          &  0.010       &  0.00          & 0.010        & N/A    \\
Party Invitations  & 80          &  0.020       &  0.01          & 0.030        & N/A    \\
Party Invitations  & 160         &  0.050       &  0.02          & 0.050        & N/A    \\
Seating		     & 9/3/3 &	  0.04    &       0.03           & 0.01         & 0.03  \\
Seating              & 16/4/4      &  11.40       &  3.72          & 0.330      & 4.337 \\
Employee Raise       & 15/5       &  0.57        &  0.87          & 0.140       & 2.750 \\
Employee Raise       & 21/15      &  2.88        &  1.75          & 1.770       & 6.235 \\
Employee Raise       & 25/20      &  3.42        &  8.38          & 5.20        & 3.95   \\
NM1                   & 125        &  1.10        &  0.07          & 1.00       & N/A   \\
NM1                   & 150        &  1.60        &  0.18          & 1.30       & N/A   \\
NM2                   & 125        &  1.44        &  0.23          & 0.80       & N/A  \\
NM2                   & 150        &  2.08        &  0.34          & 1.28       & N/A  \\
\hline
\end{tabular}}
\end{center}
\caption{Computing Answer Sets of Benchmarks with Aggregates\label{tab1}} \end{table}

\section{An Alternative Semantical Characterization} \label{sec-alternative}

The main advantage of 
the previously introduced definition of \CASP-answer sets is 
its simplicity, which allows
an easy computation of answer sets of programs with aggregates
using currently available answer set solvers.
Following this approach, all we need to do
to compute answer sets of a program $P$  
is to compute its unfolded program $unfolding(P)$ 
and then use an answer set solver to compute the answer sets of 
$unfolding(P)$. 
One disadvantage of this method lies in the fact that the size of the program 
$unfolding(P)$ can be exponential in the size of $P$---which could
potentially become unmanageable.
Theoretically, this is not a surprise, as the problem of determining 
the existence of answer sets for propositional programs with 
aggregates can be very complex,  depending on the types of aggregates (i.e.,
aggregate functions and comparison predicates) present in the 
program  (see \cite{SonP05} and Chapter 6 in \cite{Pelov04} for a thorough
discussion of these issues). 

In what follows, we present an alternative characterization of
the semantics for programs with 
aggregates, whose underlying principle is still the unfolding 
mechanism. This new characterization
allows us to compute the  answer sets of a program by 
using a generate-and-test procedure. 
The key difference is that the unfolding is now performed 
with respect to a \emph{given interpretation}.

\subsection{Unfolding with respect to an Interpretation}
\label{subsec-alternative}

Let us start by specializing the notion of solution of
an aggregate to the case of a fixed interpretation.

\begin{definition}[$M$-solution] 
For a ground aggregate atom $\ell$  and an 
interpretation $M$, its $M$-solution set is
$${\cal SOLN}^*(\ell, M) = \left\{S_\ell 
\mid S_\ell \in {\cal SOLN}(\ell),
 S_{\ell}.p \subseteq M, S_{\ell}.n \cap M = \emptyset \right\}.
$$
\end{definition}
Intuitively, ${\cal SOLN}^*(\ell, M)$ is the set of solutions of $\ell$ 
which are true in $M$. 
For a solution $S_\ell \in {\cal SOLN}^*(\ell,M)$,
the unfolding of $\ell$  w.r.t. $S_\ell$ is the conjunction 
$\bigwedge_{a \in S_{\ell} . p} a$. 
We say that $\ell'$ is an unfolding of $\ell$
with respect to $M$ if $\ell'$ is an unfolding of $\ell$ with respect 
to some $S_{\ell} \in {\cal SOLN}^*(\ell, M)$. When 
${\cal SOLN}^*(\ell, M) = \emptyset$, we say that $\bot$ 
is the only unfolding of $\ell$ in $M$. We next define the unfolding of a 
program with respect to an interpretation $M$. 

\begin{definition}[Unfolding w.r.t. an Interpretation]
\label{dunfolding2}
Let $M$ be an interpretation of the program $P$. 
The unfolding of a rule $r \in ground(P)$ w.r.t. $M$ 
is a set of rules, denoted by $unfolding^*(r,M)$,  defined as follows: 
\begin{enumerate}
\item If  
	$neg(r) \cap M \ne \emptyset$,  or if
	there is a $c \in agg(r)$ 
		such that $\bot$ is the unfolding of $c$ 
		in $M$,
	then $unfolding^*(r,M) = \emptyset$;
\item If $neg(r) \cap M = \emptyset$ and, 
	for every $c \in agg(r)$ 
	$\bot$ is not the unfolding of $c$, 
	then $r' \in unfolding^*(r,M)$ if 
	\begin{enumerate}
	\item $head(r') = head(r)$
	\item there exists a sequence of aggregate solutions 
		$\langle S_c \rangle_{c \in agg(r)}$ 
		of aggregate atoms in $agg(r)$ 
		such that $S_c \in {\cal SOLN}^*(c,M)$ for every 
		$c \in agg(r)$ and 
		$pos(r') = pos(r) \cup \bigcup_{c \in agg(r)} S_c.p$.
	\end{enumerate}
\end{enumerate}
The unfolding of $P$ w.r.t. $M$, denoted by
$unfolding^*(P,M)$, is defined as follows: 
\[ unfolding^*(P,M) = \bigcup_{r \in ground(P)} unfolding^*(r,M)\]
\end{definition}
Observe that $unfolding^*(P,M)$ is a definite program. 
Similar to the definition of an answer set in \cite{GL88}, we define 
answer sets as follows. 
\begin{definition} \label{alternatedef}
$M$ is an \CASP-answer set of $P$ iff 
$M$ is an answer set of $unfolding^*(P, M)$.
\end{definition}
In the next example, we illustrate the above definitions. 

\begin{example}
Consider the program $P_1$ (Example \ref{exp1}) and consider the
interpretation $M=\{p(a),p(b)\}$. Let $\ell$ be the aggregate atom
$\textnormal{\sc Count}(\{X \mid p(X)\}) > 0$. We have that
\[{\cal SOLN}^*(\ell,M) = \{ \langle \{p(a)\},\emptyset\rangle, 
			\langle \{p(b)\},\emptyset\rangle,
			\langle \{p(a),p(b)\},\emptyset\rangle \}\]
The $unfolding^*(P_1,M)$ is:
\[ \begin{array}{lclclclclcl}
  p(a)& \leftarrow &  p(a)  & \hspace{2cm} &
  p(a)& \leftarrow &  p(b)  \\
  p(a)& \leftarrow &  p(a), \:p(b) &&
  p(b)& \leftarrow &  & 
\end{array}
\]
Observe that $M$ is indeed an answer set of $unfolding^*(P_1,M)$.
\hfill $\Box$
\end{example}

\begin{example}
Consider the program $P_2$ from Example \ref{exp2}, and let us consider
$M = \{p(1),p(2),p(3),p(5),q\}$. Observe that, if we consider the
aggregate atom $\ell$ of the form $ \textnormal{\sc Sum}(\{X \mid p(X)\}) > 10$ then
$${\cal SOLN}^*(\ell,M) = \{ \langle \{p(1),p(2),p(3),p(5)\},\emptyset\rangle\}$$
Then, $unfolding^*(P_2,M)$ is:
\[
\begin{array}{lclclcl}
 p(1) & \leftarrow &  & \hspace{5cm} &  p(2) & \leftarrow &\\
 p(3) & \leftarrow &  & & p(5) & \leftarrow &   q\\
  q&  \leftarrow&   \multicolumn{5}{l}{p(1), \:p(2),\: p(3),\: p(5)} 
\end{array}
\]
This program has the unique answer set $\{p(1),p(2),p(3)\}$ which is
different from $M$; thus $M$ is not an answer set of $P_2$ according 
to Definition \ref{alternatedef}. \hfill $\Box$
\end{example}

\noindent
The next theorem proves that this new definition is equivalent 
to the one in Section \ref{semantics}. 
\begin{theorem}
For any \CASP\ program $P$, an interpretation $M$ 
of $P$ is an answer set of $unfolding(P)$ iff $M$
is an answer set of $unfolding^*(P,M)$.
\end{theorem}
\noindent {\bf Proof.}
Let $R = unfolding^*(P,M)$ and $Q = (unfolding(P))^M$. 
We have that $R$ and $Q$ are definite programs.
We will prove by induction on $k$ 
that if $M$ is an answer set of $Q$
then $T_Q \uparrow k = T_R \uparrow k$ for every $k \ge 0$.\footnote{$T_R$
denotes the traditional immediate consequence operator and
$T_R\uparrow k$ is the $k^{th}$ upward iteration of $T_R$.} 
The equation holds trivially for $k=0$. Let us consider 
the case for $k > 0$, assuming that $T_Q \uparrow l = T_R \uparrow l$
for $0 \le l < k$. 
\begin{itemize}
\item Consider $p \in T_Q \uparrow k$.
This means that there exists some rule $r' \in Q$ such that 
$head(r') = p$ and $body(r') \subseteq T_Q \uparrow k-1$. From 
the definition of the Gelfond-Lifschitz reduction and 
the definition of the unfolded program, we can conclude 
that there exists some rule 
$r \in ground(P)$ and a sequence of aggregate solutions
$\langle S_c \rangle_{c \in agg(r)}$ for the aggregate atoms in $body(r)$ 
such that $pos(r') = pos(r) \cup \bigcup_{c \in agg(r)} S_c.p$, and 
$(neg(r) \cup \bigcup_{c \in agg(r)} S_c.n) \cap M = \emptyset$.
In other words, $r'$ is the Gelfond-Lifschitz reduction with  
respect to $M$ of the 
unfolding of $r$ with respect to $\langle S_c \rangle_{c \in agg(r)}$.
These conditions imply that $r' \in R$. Together with the 
inductive hypothesis, we can conclude that $p \in T_R \uparrow k$.

\item Consider $p \in T_R \uparrow k$. Thus, 
there exists some rule $r' \in R$ such that 
$head(r') = p$ and $body(r') \subseteq T_R \uparrow k-1$. From 
the definition of $R$, we can conclude 
that there exists some rule 
$r \in ground(P)$ and a sequence of aggregate solutions
$\langle S_c \rangle_{c \in agg(r)}$ for the aggregate atoms in $body(r)$ 
such that $pos(r') = pos(r) \cup \bigcup_{c \in agg(r)} S_c.p$, and 
$(neg(r) \cup \bigcup_{c \in agg(r)} S_c.n) \cap M = \emptyset$.
Thus, $r' \in Q$. Together with the 
inductive hypothesis, we can conclude that $p \in T_Q \uparrow k$.
\end{itemize}
This shows that, if $M$ is an answer set of $Q$, then $M$ is an answer set of
$R$.

Similar arguments can be used to show that if $M$ is an answer set of
$R$, $T_Q \uparrow k = T_R \uparrow k$ for every $k \ge 0$, which means 
that $M$ is an answer set of $Q$.
\hfill$\Box$

\medskip \noindent
The above theorem shows that we can compute answer sets of 
aggregate programs in the same generate--and--test order 
as in normal logic programs. Given a program $P$ and an
interpretation $M$, instead of computing the  
Gelfond-Lifschitz's reduct $P^M$ we compute the
$unfolding^*(P,M)$. This method of computation might 
yield better performance but requires modifications 
of the answer set solver. 

Another advantage of this alternative characterization is
its suitability to handle aggregate atoms as heads of program
rules, as discussed next.

\subsection{Aggregates in the Head of Rules} 
\label{heads}

As in most earlier proposals, with the exception of the weight constraints
of {\sc Smodels} \cite{weight}, logic programs with 
abstract constraint atoms \cite{trus},  and answer sets for propositional 
theories \cite{ferraris}, 
the language discussed in Section \ref{sec2} does 
not allow aggregate atoms as facts (or as head of a rule).
To motivate the need for aggregate atoms as rule heads, let us
consider the following 
example. 

\begin{example}
Let us have a set of three students who have taken an exam,
and let us assume that at least two got 'A'. This can be encoded 
as the {\sc Smodels} program with the set of facts about 
students and the weight constraint 
\[
2 \{gotA(X) : student(X)\}
\]
If aggregate atoms were allowed in the head, we could encode 
this problem as the following \CASP\ program 
\[
\textnormal{\sc Count}(\{X \mid gotA(X)\}) \ge 2
\]
along with a constraint stating that if $gotA(X)$ is true 
then $student(X)$ must be true as well---which can be encoded 
using the constraint 
$$\bot \leftarrow gotA(X), \naf student(X)$$
This program should have four answer sets, each representing
a possible grade distribution, in which either one of the 
students does not receive the 'A' grade or all the three
students receive 'A'. \hfill $\Box$ 
\end{example}

The above example suggests that aggregate atoms in the head of a rule are convenient
for certain knowledge representation tasks. We will now consider logic 
programs with aggregate atoms in which each rule is an expression of the form
\begin{equation} \label{agg-rule-2}
D \leftarrow C_1,\ldots,C_m,A_1,\ldots,A_n,\naf B_1, \dots, \naf B_k
\end{equation}
where $A_1, \dots, A_n, B_1, \dots, B_k$ are ASP-atoms, 
$C_1, \dots, C_m$ are aggregate atoms ($m\geq 0$, $n\geq 0$, $k\geq 0$), 
and $D$ can be either an ASP-atom or an aggregate atom. 
An \CASP\ program is now a collection of rules of the above form.
The notion of a model can be straightforwardly generalized 
to program with aggregates in the head. It is omitted here for brevity. 

As it turns out, the semantics presented in the previous subsection
can be easily extended to allow for aggregate atoms in the head 
of rules. It only requires an additional step, in order to convert programs with 
aggregates in the head to programs without aggregates in the head. 
To achieve that, we introduce the following notation. 

\begin{definition}
Let $P$ be a program with aggregates in the head,
$M$ be an interpretation of $P$, and 
$r$ be one of the rules in $ground(P)$ such that $head(r)$ is an aggregate atom.
We define $r^\bot = \{\bot \leftarrow body(r)\}$
and $r^M = \{p \leftarrow body(r) \mid p \in {\cal H}(head(r)) \cap M\}$.
\end{definition}

\begin{definition} \label{aggheadfree}
Let $P$ be a program with aggregates in the head and let
$M$ be an interpretation of $P$. The 
 {\em aggregate-free head reduct} of $P$ with 
respect to $M$, denoted by $P(M)$, is the program obtained 
from $P$ by replacing each rule $r \in ground(P)$ whose head
is an aggregate atom with 
\begin{itemize}
\item[(a)] $r^\bot$ if ${\cal SOLN}^*(head(r),M) = \emptyset$; 
or 

\item[(b)] $r^M$ if  ${\cal SOLN}^*(head(r),M) \ne \emptyset$. 
\end{itemize}
\end{definition}
For each rule $r$, whose head is an aggregate atom, we first check 
whether $head(r)$ is satisfied by $M$ (i.e.,
${\cal SOLN}^*(head(r),M) = \emptyset$ by Observation \ref{obs0}). If it is not
satisfied, then this means that we intend the rule's body to not
be satisfied---and we encode this with a rule of the type $r^{\bot}$. Otherwise,
$M$ provides us with a solution of the aggregate atom $head(r)$---i.e.,
$\langle M\cap {\cal H}(head(r)), \; {\cal H}(head(r))\setminus M\rangle$\/---and
we intend to use this rule to ``support'' such solution; in particular, the
rules in $r^M$ provides support for all the elements in
$M\cap {\cal H}(head(r))$. 
We are now ready to define the notion of answer sets for program
with aggregates in the head. 
\begin{definition} \label{def-general}
A set of atoms $M$ is an \CASP-answer set of $P$ iff
$M$ is an answer set of $unfolding^*(P(M),M)$.
\end{definition}

Observe that, because 
of aggregates in the head, an \CASP-answer set might be
non minimal. Nevertheless, the following holds. 

\begin{observation} \label{asmodel}
Every \CASP-answer set of a program $P$
is a model of $P$.
\end{observation}

\begin{example}
Consider the program $P_5$:
\[
\begin{array}{rclcl}
student(a) & \leftarrow &  \hspace{1cm} \\
student(b) & \leftarrow &  \hspace{1cm} \\
student(c) & \leftarrow &  \\
\textnormal{\sc Count}(\{X \mid gotA(X)\}) \ge 2 & \leftarrow &  \\
\bot & \leftarrow & gotA(X), \naf student(X) 
\end{array}
\]

Let us compute some answer sets of $P_5$.  
Let $\ell$ denote the aggregate atom
$\textnormal{\sc Count}(\{X \mid gotA(X)\})\geq 2$.

\begin{itemize}
\item 
Let $M_1 = \{student(a), student(b), student(c), gotA(a)\}$. 
We can check that $\ell$ is not satisfied by $M_1$, and hence, 
the unfolding of the fourth rule of 
$P_5$ is the set of rules $\{\bot\}$, i.e., $unfolding^*(P_5(M_1),M_1)$
is the following program:
\[
\begin{array}{rclcl}
student(a)& \leftarrow & \hspace{1cm}& \\
student(b)& \leftarrow &\hspace{1cm} & \\
student(c)& \leftarrow & \\
\bot  \\
\bot & \leftarrow & gotA(X), \naf student(X)
\end{array}
\]
This program does not have any answer set. 
Thus, $M_1$ is not an \CASP-answer set of $P_5$ (according to Definition \ref{def-general}).

\item Consider 
$M_2 = \{student(a), student(b), student(c), gotA(a), gotA(b)\}.$
We have that $\ell$ is satisfied by $M_2$. Hence, $unfolding^*(P_5(M_2),M_2)$ 
is obtained from $P_5$ by replacing its fourth rule with the two rules 
\[
\begin{array}{rclcl}
gotA(a) & \leftarrow &  \hspace{1cm} \\
gotA(b) & \leftarrow \\
\end{array}
\]
$unfolding^*(P_5(M_2),M_2)$ has $M_2$ as an answer set. Therefore, $M_2$ is an 
\CASP-answer set of $P_5$.
\end{itemize}
Similar to the second item, we can show that 
\[
\begin{array}{lll}
M_3 & = & \{student(a), student(b), student(c), gotA(a), gotA(c)\} \\
M_4 & = & \{student(a), student(b), student(c), gotA(b), gotA(c)\} \\
M_5 & = & \{student(a), student(b), student(c), gotA(a), gotA(b), gotA(c)\} \\
\end{array}
\]
are answer sets of $P_5$.\hfill $\Box$
\end{example}

\section{Related Work}
\label{related-work}

In this section,  we relate our definition of \CASP-answer sets 
to several formulations of aggregates proposed in the literature. 
We begin with a comparison of our unfolding approach 
 with the two most recently proposed 
semantics for LP with aggregates, i.e., 
the \emph{ultimate stable model semantics} 
\cite{PelovDB04,PelovDB03,Pelov04}, the 
\emph{minimal answer set semantics} 
\cite{faberLP04,ferraris}, and the semantics for 
abstract constraint atoms \cite{trus}. 
We then relate our work to earlier proposals, 
such as perfect models of
aggregate-stratified programs (e.g., \cite{MumickPR90}), the
fixpoint answer set semantics of aggregate-monotonic programs \cite{KempS91},
and the semantics of programs with weight constraints \cite{weight}. Finally,
we briefly discuss the relation of \CASP-answer sets to other proposals. 

\subsection{Pelov's Approximation Semantics for Logic Program with Aggregates}
\label{pelov}

The doctoral thesis of Pelov \cite{Pelov04} contains a nice 
generalization of several semantics of logic programs to the case of logic 
programs with aggregates. The key idea in his work is the use of 
approximation theory in defining several semantics for logic programs
with aggregates (e.g., two-valued semantics, ultimate three-valued
stable semantics, three-valued stable model semantics). In particular,
in \cite{Pelov04}, the author describes a fixpoint operator, called
$\Phi^{appr}_P$, operating on 3-valued interpretations and parameterized
by the choice of approximating aggregates. The results presented in
\cite{SonP05} allow us to conclude the following result.

\begin{proposition}
Given a program with aggregates $P$, $M$ is a \CASP\ answer set of $P$
if and only if $M$ is the least fixpoint of
$\Phi^{aggr,1}_P(\cdot,M)$, where $\Phi^{aggr,1}_P$ denotes the first
	component of $\Phi^{aggr}_P$.
\end{proposition}
The work of Pelov includes also a translation of logic programs 
with aggregates to normal logic programs, denoted by $tr$, which was 
first given in \cite{PelovDB03} and then in \cite{Pelov04}. The 
translation in \cite{PelovDB03} (independently developed)
and the unfolding proposed in Section \ref{sec2} have
strong similarities\footnote{We 
	would like to thank a reviewer of an earlier version of this paper 
  who provided us with the pointers to these works.}.
 For the 
completeness of the paper, we will review the basics of the translation 
of \cite{PelovDB03}, expressed using our notation.
Given a logic program with aggregates $P$, $tr(P)$ denotes the 
normal logic program obtained after the translation. The translation 
begins with the translation of each aggregate atom
$\ell = {\cal P}(s)$ 
into a disjunction $tr(\ell) = \bigvee F^{{\cal H}(s)}_{(s_1,s_2)}$ where 
${\cal H}(s)$ is the set of atoms of $p$---the predicate of $s$---in 
${\cal B}_P$, 
$(s_1,s_2)$ belongs to an index set, 
 $s_1 \subseteq s_2 \subseteq {\cal H}(s)$, and 
each $F^{{\cal H}(s)}_{(s_1,s_2)}$ is a conjunction of the form 
\[
\bigwedge_{l \in s_1} l \wedge \bigwedge_{l \in s \setminus s_2} \naf l 
\]
The construction of $tr(\ell)$ considers only pairs $(s_1,s_2)$
satisfying the following condition:  
every interpretation $I$ such that  $s_1 \subseteq I$ 
and $s\setminus s_2 \cap I = \emptyset$ also satisfies $\ell$.
$tr(P)$ is then created by rewriting rules with disjunction 
in the body by a set of rules in a straightforward way. For example, 
the rule 
\[
a \leftarrow (b \vee c), d
\]
is replaced by the two rules 
\[
\begin{array}{lcl} 
a &\leftarrow& b, d \\
a &\leftarrow& c, d\\
\end{array}
\]
We can prove a lemma that connects $unfolding(P)$ and $tr(P)$.
\begin{lemma} \label{tr1}
For every aggregate atom $\ell = {\cal P}(s)$,
$S$ is a solution of $\ell$ if and only if 
$F^{{\cal H}(s)}_{(S.p,S.p \cup ({\cal H}(s) \setminus S.n))}$ is a disjunct in $tr(\ell)$.
\end{lemma}

\noindent{\bf Proof.}
The result is a trivial consequence of  the definition of a solution and 
the definition of $tr(\ell)$.
\hfill$\Box$

\medskip \noindent
This lemma allows us to prove the following relationship between 
$unfolding(P)$ and $tr(P)$.
\begin{corollary}
For every program $P$, 
$A$ is an \CASP-answer set of $P$ if and only if $A$ is an exact  
stable model of $P$ with respect to \cite{PelovDB04}.
\end{corollary}
\noindent {\bf Proof.}
The result is a trivial consequence of the
fact that  $unfolding(P) = tr(P)$ 
and $tr(P)$ has the same set of partial stable models 
as $P$ \cite{PelovDB03}. 
\hfill $\Box$

\subsection{\CASP-Answer Sets and Minimality Condition}
\label{mini}

In this subsection, we investigate the relationship between 
\CASP-answer sets and the notion of answer set 
defined by {\em Faber et al.} 
in \cite{faberLP04}. The notion of answer set proposed
 in \cite{faberLP04}
is based on a new notion of reduct, defined as follows. 
Given a program $P$ and a set of 
atoms $S$, the {\em reduct of P with respect to S}, denoted by 
${^S}P$, is obtained by removing from 
$ground(P)$ those rules whose body is not satisfied by $S$.
In other words,
\[
{^S}P = \{r \mid r \in ground(P), S \models body(r)\}.
\]
The novelty of this reduct is that it 
\emph{does not} remove aggregate atoms
and negation-as-failure literals satisfied by $S$. 

\begin{definition} [FLP-answer set, \cite{faberLP04}]
\label{d-faberLP04}
For a program $P$, $S$ is a {\em FLP-answer set} of $P$ 
iff it is a minimal model of ${^S}{P}$.  
\end{definition}
Observe that the definition of an answer set 
in this approach \emph{explicitly} requires answer sets to be
minimal, thus requiring the ability to determine
minimal models of a program with aggregates. 
In the following propositions, we
will show that \CASP-answer 
sets of a program $P$ are FLP-answer sets 
and that FLP-answer sets of $P$ are minimal models 
of $unfolding(P)$\/, but not necessary \CASP-answer sets.

\begin{theorem}
\label{th2}
Let  $P$ be a program with aggregates. If $M$ is an \CASP-answer set, 
then $M$ is a FLP-answer set of $P$.
If $M$ is a FLP-answer set of $P$ then
$M$ is a minimal model of $unfolding(P)$.
\end{theorem}
\noindent
{\bf Proof.} 
\begin{itemize}
\item Let $Q = unfolding(P)$. Since $M$ is an \CASP-answer set, 
we have that $M$ is an answer set of $Q$.
Lemma \ref{lem2} shows that $M$ is a 
model of $ground(P)$ and hence is a model of $R = {^M}(ground(P))$. 

Let us assume that $M$ is not a minimal model of $R$. 
This means that there exists $M' \subsetneq M$ such that $M'$ is a model 
of $^M(P)$. 

We will show that $M'$ is a model of $Q' = Q^M$ where $Q^M$ is the 
result of the Gelfond-Lifschitz transformation of the program $Q$ 
with respect to $M$. 

Consider a rule $r_2 \in Q'$ such that $M' \models body(r_2)$,
i.e., $pos(r_2) \subseteq M'$. From 
the definition of the Gelfond-Lifschitz transformation, 
we conclude that there exists some $r' \in Q$ such that 
$pos(r') = pos(r_2)$ and $neg(r') \cap M = \emptyset$. 
This implies that there is a rule $r \in ground(P)$ 
and a sequence of solutions $\langle S_c \rangle_{c \in agg(r)}$
of aggregates in $r$ 
such that $r'$ is the unfolding of $r$ with respect to 
$\langle S_c \rangle_{c \in agg(r)}$ and for every $c \in agg(r)$,
$S_c.p \subseteq M'$ and 
$S_c.n \cap M = \emptyset$. Since $M' \subseteq M$, 
we can conclude that $M \models body(r)$, i.e., 
$r \in R$. Furthermore, $M' \models body(r)$
because $pos(r) \subseteq pos(r') = pos(r_2) \subseteq M'$,
$neg(r) \subseteq neg(r')$ and $neg(r') \cap M = \emptyset$, 
and for every $c \in agg(r)$, $S_c.p \subseteq M'$ and 
$S_c.n \cap M' = \emptyset$. Since $M'$ is
a model of $R$, we have that $head(r) \in M'$. 
Since $head(r_2) = head(r') = head(r)$, we have that $M'$ satisfies 
$r_2$. This holds for every rule of $Q'$. Thus,
$M'$ is a model of $Q$. This contradicts the 
fact that $M$ is an answer set of $Q$.

\item Let $M$ be a FLP-answer set of $P$. Clearly, $M$ is a model of 
$ground(P)$ and hence of $unfolding(P)$. If $M$ is not a minimal model
of $unfolding(P)$, there exists some $M' \subsetneq M$ which is a model 
of $unfolding(P)$. Lemma \ref{lem1} implies that $M'$ is a model of
$ground(P)$ and hence is a model of ${^M}P$. This is a contradiction 
with the assumption that $M$ is a FLP-answer set of $P$. Thus,
we can conclude that $M$ is a minimal model of $unfolding(P)$.
\end{itemize}
\qed

\medskip
\noindent
The next example shows that FLP-answer sets 
might not be  \CASP-answer 
sets.\footnote{
  We would like to thank an anonymous reviewer of an earlier version of this 
  paper who suggested this example. 
}

\begin{example}\label{p5}
Consider the program $P_6$ where 
\[
\begin{array}{lll}
p(1)& \leftarrow & \textnormal{\sc Sum}(\{X \mid p(X)\}) \ge 0 \\
  p(1)& \leftarrow & p(-1)  \\
 p(-1)& \leftarrow & p(1) \\
\end{array}
\]
The interpretation $M = \{p(1), p(-1)\}$ is a FLP-answer set of $P_6$. 
We will show next that 
$P_6$ does not have an answer set according to our definition.
It is possible to show\footnote{We follow the common practice
that the sum of an empty set is equal to $0$.} that the aggregate atom
$\textnormal{\sc Sum}(\{X \mid p(X)\}) \geq 0$ has the 
following solutions with respect to ${\cal B}_P = \{p(1),p(-1)\}$: 
$\langle \emptyset, \{p(-1)\} \rangle$,
$\langle \emptyset, \{p(1),p(-1)\} \rangle$,
$\langle \{p(1)\}, \{p(-1)\} \rangle$,
$\langle \{p(1)\}, \emptyset \rangle$, and
$\langle \{p(1),p(-1)\}, \emptyset \rangle$. The unfolding of $P_6$,
$unfolding(P_6)$,  
consists of the following rules:
\[
\begin{array}{llll}
p(1) & \leftarrow & \naf p(-1)\\
p(1)& \leftarrow & \naf p(1), \naf p(-1) \\
p(1)& \leftarrow & p(1), \naf p(-1) \\
p(1)& \leftarrow & p(1) \\
p(1)& \leftarrow & p(1), p(-1) \\
p(1)& \leftarrow & p(-1)  \\
p(-1)& \leftarrow & p(1) \\
\end{array}
\]  
It is easy to see that $unfolding(P_6)$ does not have answer sets. 
Thus, $P_6$ does not have  \CASP-answer sets.\hfill $\Box$
\end{example}

\begin{remark} 
\label{rem1}
If we replace in $P_6$ the rule
$p(1) \leftarrow \textnormal{\sc Sum}(\{X \mid p(X)\}) \ge 0$ 
with the intuitively equivalent {\sc Smodels} weight constraint 
rule $$p(1) \leftarrow 0 [p(1) = 1, p(-1)=-1].$$ 
we obtain a program that does not have answer sets 
in {\sc Smodels}.
\end{remark}

The above example shows that our characterization of 
programs with aggregates differs from the proposal in
\cite{faberLP04}. 
Apart from the lack of support
for aggregates in the heads of rules, the semantics
of~\cite{faberLP04} might accept answer sets that
are not \CASP-answer sets. 
Observe that the two semantical characterizations 
coincide for large classes of programs (e.g., for
programs that have only monotone aggregates).

\subsection{Logic Programs with Abstract Constraint Atoms}

A very general semantic characterization of programs with 
aggregates has been proposed by Marek 
and Truszczy\'{n}ski in~\cite{trus}. The framework offers a model 
where general aggregates can be employed both in the body and in the head
of rules. The authors introduce the notion of abstract constraint atom, $(X,C)$,
where
$X$ is a set of atoms (the domain of the aggregate) and
 $C$ is a subset of $2^X$ (the solutions of the aggregate). For 
an abstract constraint atom $A=( X,C )$, we will denote
$X$ with $A_D$ and $C$ with $A_C$. 
In \cite{trus}, the
focus is only on \emph{monotone constraints}, i.e., constraints $(X,C)$, where
if $Y \in C$ then all supersets of $Y$ are also in $C$.

A program with monotone constraints is a set of rules of the form
\[ B_0 \leftarrow B_1, \dots, B_n, not\:B_{n+1}, \dots, not\:B_{n+m} \]
where each $B_i$ ($i \geq 0$) is an abstract constraint atom. 
Abusing the notation, for a rule $r$ of the above form, 
we use $head(r)$, $pos(r)$, $neg(r)$, 
and $body(r)$ to denote 
$B_0$, $\{B_1, \dots, B_n\}$, $\{B_{n+1}, \dots, B_{n+m}\}$,
and $\{B_1, \dots, B_n, not\:B_{n+1}, \dots, not\:B_{n+m}\}$, respectively.
The semantics of this language is developed as a generalization of 
answer set semantics for normal logic programs. To make the paper 
self-contained, we briefly review the notion of a stable model for 
a program with monotone constraints. 

An interpretation $M$ satisfies $A = (X,C)$, denoted by $M \models A$, 
if $X\cap M \in C$ (or $A_D\cap M \in A_C$). $M \models \naf A$ if $M \not\models A$. 
For a set of literals $S$, $M \models S$ if $M \models B$ for each $B \in S$. 
For a program with monotone constraints $P$, $hset(P)$ denotes the set 
$\cup_{r \in P} head(r)_d$. Given a set of atoms $S$, 
a rule $r$ is {\em applicable} in $S$ if $S \models body(r)$. 
The set of applicable rules in $S$ is denoted by $P(S)$. 
A set $S'$ is {\em nondeterministically one-step provable}
from $S$ by means of $P$ if $S' \subseteq hset(P(S))$ and 
$S' \models head(r)$ for every $r \in P(S)$. 
The {\em nondeterministic one-step provability operator} $T^{nd}_P$ 
is a function from $2^{\cal A}$ to $2^{2^{\cal A}}$, where 
$\cal A$ denotes the Herbrand base of $P$, such that 
for every $S \subseteq {\cal A}$, $T^{nd}_P(S)$ consists of all sets $S'$
that are nondeterministically one-step provable from $S$ by means of $P$. 
A sequence $t=(X_n)_{n=0,1,2,\dots}$ is called a 
\emph{P-computation} if  $X_0=\emptyset$ and for every non-negative 
integer $n$, 
\begin{itemize}
\item [({\em i})] $X_n \subseteq X_{n+1}$, and 
\item [({\em ii})] $X_{n+1} \in T^{nd}_P(X_n)$.
\end{itemize}
$S_t = \cup_{n=0}^\infty X_i$ is called the {\em result} of the computation $t$. 
A set of atoms $S$ is a {\em derivable model} of $P$ if there exists 
a $P$-{\em computation} $t$ such that $S = S_t$. For 
a monotone program $P$ and a set of atoms $M$, the {\em reduct} of 
$P$ with respect to $M$, denoted by $P^M$, is obtained from $P$ by 
({\em i}) removing from $P$ every rule containing in the body a literal 
$\naf A$ such that $M \models A$; and ({\em ii}) 
 removing all literals of the form $\naf A$ from the remaining rules.
A set of atoms $M $ is an {\em stable model} of a monotone program $P$ if $M$ is 
a derivable model of the reduct $P^M$. 

Observe that each aggregate atom $\ell$ in our notation can be represented by an 
abstract constraint atom $({\cal H}(\ell), C_{\ell})$, where 
$C_\ell = \{S \mid S \subseteq {\cal H}(\ell), S \models \ell\}$. 
Furthermore, an atom $a$ can be represented as 
an abstract constraint atom $(\{a\},\{\{a\}\})$. 
Thus, each 
program $P$, as a set of rules of the form (\ref{agg-rule-2}), could be viewed 
as a program with abstract constraint atoms $P_A$, where $P_A$ is obtained 
from $P$ by replacing every occurrence of an aggregate atom $\ell$ 
or an atom $a$ in $P$ 
with $( {\cal H}(\ell), C_{\ell} )$ or 
$(\{a\},\{\{a\}\})$ respectively. 
The monotonicity of an abstract constraint atom
implies the following:
\begin{observation} \label{obs2}
Let $\ell$ be an aggregate atom and $M$ be a set of atoms such 
that $C_\ell({\cal H}(\ell))$  is a monotone constraint and 
$M \models C_\ell({\cal H}(\ell))$. Then, 
$\langle M \cap {\cal H}(\ell), \emptyset \rangle$ is a solution of 
$\ell$.  
\end{observation}
Using this observation, we can related the notions 
of \CASP-answer set and of stable models for programs with monotone 
atoms as follows.

\begin{theorem}
Let $P$ be a program with monotone aggregates. $M$ is  an
\CASP\ answer set of $P$ iff $M$ is a stable model of $P_A$ 
according to \cite{trus}.
\end{theorem}

\noindent
{\bf Proof.} 
For each rule $r \in P$, let $r_A$ be the rule 
in $P_A$ which is obtained from $r$ by the translation from $P$ to $P_A$. 

\begin{enumerate}

\item[``$\Longleftarrow$''] Let us assume $M$ is a stable model of $P_A$
according to~\cite{trus}. A result in \cite{trus} shows that 
$M = \bigcup_0^\infty X_i$ where 
\[\begin{array}{lcl}
	X_0 & = & \emptyset \\
	X_{i+1} & =& M \cap hset(P(X_i))
  \end{array}
\]
We will show that $M$ is a \CASP\ answer set of $P$ by proving 
that $lfp(T_Q) = M$ where $Q = unfolding^*(P',M)$ and $P'$ the 
aggregate-free head reduct of $P$ with respect to $M$ 
(Definition \ref{aggheadfree}). 

Let us start by showing that $X_i \subseteq T_Q\uparrow i$ 
for $i \geq 0$, using induction on $i$. The result is obvious 
for $i=0$. Let us assume the result to hold for $i \leq k$ and 
let us consider the case of $X_{k+1}$. By the definition of 
$X_i$'s, $p\in X_{k+1}$ implies that $p \in M \cap hset(P_A(X_{k}))$. 
This means that there is a rule 
\begin{equation}\label{eqq1}
A \leftarrow B_1, \dots, B_n \in P_A^M
\end{equation}
such that $X_{k}\models B_i$ for $i=1,\dots,n$, $A_D \cap M \in A_C$,
and $p \in A_D \cap M$. From $X_{k}\cap (B_i)_D \in (B_i)_C$, 
$X_{k}\cap (B_i)_D \subseteq T_Q\uparrow k \cap (B_i)_D$, and the
monotonicity of $B_i$, we have 
that $T_Q\uparrow k \cap (B_i)_D \in (B_i)_C$. From 
Observation~\ref{obs2} and the monotonicity of the aggregates, 
we can infer that there is a rule in $Q$ with $head(r) = p$ and
$T_Q\uparrow k \models body(r)$. Thus, $p \in T_Q \uparrow (k+1)$.
The inductive step is proved. 
This allows us to conclude that $M \subseteq lfp(T_Q)$. 

On the other hand, we can easily show that $M$ is a model of
$Q$, thus $lfp(T_Q) \subseteq M$. This allows us to conclude
that $M = lfp(T_Q)$. Together with $M \subseteq lfp(T_Q)$,
we have that $M$ is an \CASP\ answer set of $P$. 

\item[``$\Longrightarrow$''] Let $M$ be an \CASP\ answer set of $P$ and 
$Q = unfolding^*(P',M)$ where 
$P'$ is the  aggregate-free head reduct of $P$ with 
respect to $M$. Thus, $M=lfp(T_Q)$. 

We will prove that $M$ is a stable model of $P$ by showing that 
the sequence $X_i = T_Q\uparrow i$, for $i \geq 0$, is a P-computation. 
Obviously, we have that ({\em i.}) $X_0 = \emptyset$ and 
({\em ii}) $X_i  \subseteq X_{i+1}$ for $i\geq 0$. It remains to be shown 
that ({\em iii}) $X_{i+1} \in T_P^{nd}(X_i)$ for $i \geq 0$. 

In order to prove the property (\emph{iii}) we need to show that
({\em iv}) $X_{i+1} \subseteq hset(P_A(X_i))$ and
({\em v}) $X_{i+1} \models head(r)$ for each $r_A$ in $P_A(X_i)$.

Let us consider $p \in X_{i+1}=T_Q \uparrow i+1$. This means 
that there exists some rule $r'$ in $Q$ such that $body(r') \subseteq X_i$ and 
$head(r') = p$. Let $r$ be the rule in $P$ such that 
$r'$ is obtained from $r$ (as specified in Definitions 
\ref{aggheadfree}-\ref{dunfolding2}). This implies   
$neg(r) \cap M =\emptyset$, $pos(r) \subseteq X_i$, and
$X_i \models c$ for every $c \in agg(r)$. From the monotonicity
of aggregates in $P$, we can conclude that $r_A \in P_A(X_i)$ and 
$p \in head(r_A)_D$. This holds for every $p \in X_{i+1}$.
Hence, we have that $X_{i+1} \subseteq hset(P_A(X_i))$, i.e.,
({\em iv}) is proved. 

Now let us consider a rule $r_A \in P_A(X_i)$. 
This implies that the rule $r$, from which $r_A$ is obtained, satisfies 
that $neg(r) \cap M =\emptyset$, $pos(r) \subseteq X_i$, and
$X_i \models c$ for every $c \in agg(r)$. Again, the  
monotonicity of aggregates in $P$ implies that 
$M \models c$ for every $c \in agg(r)$. By the definition of $Q$, 
we have that for each $p \in head(r_A)_D \cap M$, there exists 
a rule $r_p \in Q$ such  that $p = head(r_p)$ and 
$body(r_p) \subseteq X_i$. As such, $head(r_A)_D \cap M \subseteq X_{i+1}$.
This means that $X_{i+1} \models head(r_A)$, i.e., ({\em v}) is proved.
\qed
\end{enumerate}

\subsection{Answer Sets for Propositional Theories}

The proposal of Ferraris \cite{ferraris} applies a novel notion of
reduct and answer sets, developed for propositional theories, to the
case of aggregates containing arbitrary formulae. The intuition behind
the notion of satisfaction of an aggregate relies on translating 
aggregates to propositional formulae that guarantee that all cases
where the aggregate is false are ruled out. In particular, for an aggregate 
of the form $F(\{\alpha_1=w_1,\dots,\alpha_k=w_k\}) \odot R$, where
$\alpha_i$ are propositional formulae, $w_j$ and $R$ are real numbers, $F$ 
is a function from multisets of real numbers to $\mathbb{R}\cup\{+\infty,-\infty\}$,
and $\odot$ is a relational operator (e.g., $\leq$, $\neq$), 
the transformation leads to the propositional
formula:
\[
\bigwedge_{\footnotesize\begin{array}{c}
		I\subseteq\{1,\dots,k\}\\
		F(\{w_i\:|\: i\in I\})\not{\!\odot}R
	   \end{array}}
	\left(
		\left(
			\bigwedge_{i\in I} \alpha_i
		\right)
		\Rightarrow
		\left(
			\bigvee_{i\in \{1,\dots,k\}\setminus I}
				\alpha_i
		\right)
	\right)
\]
The results in~\cite{ferraris} show that the new notion of reduct, along
with this translation for aggregates, applied to the class of
logic programs with aggregates of \cite{faberLP04}, captures exactly the class of 
FLP-answer sets.

\subsection{Logic Programs with Weight Constraints}

Let us consider the weight constraints employed by {\sc Smodels} and let us 
describe a translation method to convert them into our language with aggregates. 
We will focus on weight constraint that are used in the body of 
rules (see Sect.~\ref{heads} for aggregates in the heads of rules). 
For simplicity, we will also focus on weight 
constraints with non-negative weights (the generalization can 
be obtained through 
algebraic manipulations, as described in \cite{weight}).
A  \emph{ground} weight constraint $c$ has the form:\footnote{Note
that grounding removes {\sc Smodels}' conditional literals.}
\[
L \leq \{p_1{=}w_1,\dots,p_n{=}w_n,not\:r_1{=}v_1,\dots,not\:r_m{=}v_m\}\leq U
\]
where $p_i, r_j$ are ground atoms, and $w_i, v_j, L, U$ are numeric constants. 
$p_i$'s and $\naf r_j$'s are called literals of $c$.
$lit(c)$ denotes the set of literals of $c$.
The local weight function of a constraint $c$, $w(c)$, 
returns the weight of its literals. For example, $w(c)(p_i) = w_i$
and $w(c)(\naf r_i) = v_i$. The weight of a weight constraint $c$
in a model $S$, denoted by $W(c,S)$, is given by 
\[ 
W(c,S) = \sum_{p \in lit(c), \: p \in S } w(c)(p) +
	\sum_{not\:q \in lit(c), \: q \notin S} w(c)(not\:p).  
\]

We will now show how weight constraints in {\sc Smodels} 
can be translated into aggregates in our language. 
For each weight constraint $c$, let $agg^+_c$ and $agg^-_c$
be two new predicates which do not belong to the language of $P$.
Let $r(c)$ be the set of following rules: 
\[
\begin{array}{lll}
agg^+_c(1,w_1)  \leftarrow  p_1. & \hspace*{1cm} \cdots \hspace*{1cm} &
agg^+_c(n,w_n)  \leftarrow  p_n.  \\
agg_c^-(1,v_1) \leftarrow  r_1. & \hspace*{1cm} \cdots \hspace*{1cm} &
agg_c^-(m,v_m)  \leftarrow  r_m. \\
\end{array}
\]
Intuitively, $agg_c^+, agg_c^-$ assign
a specific weight to each literal originally present 
in the weight constraint. 
The weight constraint itself is replaced by a conjunction $\tau(c)$:
\[
\tau(c) = \left\{
\begin{array}{c}
\textnormal{\sc Sum}(\{\!\!\{X\mid\exists Y. agg_c^+(Y,X)\}\!\!\}) = S^+ \:\:\wedge\:\:  
\textnormal{\sc Sum}(\{\!\!\{X\mid\exists Y. agg_c^-(Y,X)\}\!\!\}) = S^- \:\:\wedge \\
			L \leq S^+ + \sum_{i=1}^m v_i - S^- \leq U
\end{array}\right.
\]
where {\sc Sum} is an aggregate function with its usual meaning.

Given an {\sc Smodels} program $P$, let $\tau(P)$ be the program 
obtained from $P$ by  replacing every weight constraint $c$ 
in $P$ with $\tau(c)$ and adding the set of rules $r(P)$ to $P$
where $r(P) = \bigcup_{c \textnormal{ is a weight constraint in } P} r(c)$.
For each set of atoms $S$, 
let us denote with $\hat S = S \cup T_{r(P)}(S)$.\footnote{$T_{r(P)}$ is
the immediate consequence operator of program $r(P)$.} We have that
\begin{equation}\label{hats}
{\hat S} = 
\begin{array}{lll}
S & \cup & \{agg^+_c(i,w_i) \mid c \textnormal{ is a weight constraint in } P, p_i=w_i \in c, p_i \in S\} \\
	 &   \cup & \{agg^-_c(i,v_i) \mid c \textnormal{ is a weight constraint in } P, \naf q_i=v_i \in c, q_i \in S\}. 
\end{array}
\end{equation}
This implies the following lemma. 
\begin{lemma} 
Let $S$ be a set of atoms and $c$ be a weight constraint.
For $\hat S = S \cup T_{r(P)}(S)$, 
\[
W(c,S) = 
\textnormal{\sc Sum}(\{\!\!\{X\mid\exists Y. agg_c^+(Y,X)\}\!\!\})^{\hat S} + \sum_{i=1}^m v_i - 
\textnormal{\sc Sum}(\{\!\!\{X\mid\exists Y. agg_c^-(Y,X)\}\!\!\})^{\hat S}
\]
\end{lemma}
\noindent
{\bf Proof.} Follows directly from Equation \ref{hats} and the definition of $W(c,S)$.
\hfill$\Box$

\begin{corollary} \label{weightc}
Given a set of atoms $S$ and a weight constraint $c$, $S\models c$ iff
${\hat S} \models \tau(c)$.
\end{corollary}
The next theorem relates $P$ and $\tau(P)$. 
\begin{theorem}
Let $P$ be a ground {\sc Smodels} program with weight constraints only in the 
body and with no negative literals in the weight constraints. Let 
$\tau(P)$ be its translation to aggregates. It holds that 
\begin{enumerate}
\item if $S$ is an {\sc Smodels} 
answer set of $P$ then ${\hat S}$ is an \CASP-answer set of $\tau(P)$;
\item if $\hat S$ is an \CASP-answer set of $\tau(P)$ then 
${\hat S} \cap lit(P)$ is a {\em minimal} {\sc Smodels} answer set of $P$.
\end{enumerate}
\end{theorem}
\noindent {\bf Proof.} 
Since negation-as-failure literals can be replaced by weight constraints,
without loss of generality, we can assume that $P$ is a 
positive program with weight constraints. 
Let $S$ be a set of atoms and $R$ be the {\sc Smodels}
reduct of $P$ with respect to $S$. 
Furthermore, let $Q = (unfolding(\tau(P)))^{\hat S}$. 
Using Corollary \ref{weightc}, we can prove by induction on $k$ that 
if  $S$ is an {\sc Smodels} answer set of $P$
(resp. ${\hat S}$ is an answer set of $\tau(P)$) then 
\begin{enumerate}
\item $T_Q \uparrow k \subseteq \widehat{(T_R \uparrow k)}$ for $k \ge 0$ 
\item $T_R \uparrow k \subseteq (T_Q \uparrow k) \cap lit(P)$ for $k \ge 0$
\end{enumerate}
This proves the two items of the theorem. 
%
\hfill$\Box$

\medskip \noindent
The following example, used in \cite{PelovDB04} to show that 
{\sc Smodels}-semantics for weight constraints is counter-intuitive in 
some cases, indicates that the equivalence does not hold when 
negative literals are allowed in the weight constraint. 
\begin{example}
Let us consider the {\sc Smodels} program $P_7$ 
\[
\begin{array}{l}
 p(0) \leftarrow  \{ \naf p(0)=1 \} 0\\
\end{array}
\]
According to the semantics described in \cite{weight}, we can observe that, 
for $S = \emptyset$, the reduct $P_7^S$ is $\emptyset$
making it an answer set of $P_7$. For $S = \{ p(0) \}$, the reduct $P_7^S$
is 
\[\begin{array}{l}
  p(0) \leftarrow
  \end{array}
\]
thus making $\{p(0)\}$ 
an answer set of $P_7$.

On the other hand, the intuitively equivalent program using aggregates (we
make use of the obvious extension that allows negations in the aggregate) is:
\[\begin{array}{l}
 p(0) \leftarrow \textnormal{\sc Count}(\{X\mid \naf p(X)\}) \leq 0.\\
  \end{array}
\]
The unfolding of this program is
\[
\begin{array}{l}
 p(0) \leftarrow p(0).
\end{array}
\]
which has the single answer set $\emptyset$. \hfill $\Box$
\end{example}

\subsection{Stratified Programs}\label{stra}
Various forms of stratification (e.g., lack of recursion through
aggregates) have been proposed to syntactically identify classes 
of programs that admit a unique minimal model, e.g., local stratification 
\cite{MumickPR90}, modular stratification \cite{MumickPR90}, and 
XY-stratification \cite{ZanioloAO93}. Efficient evaluation strategies 
for some of these classes have been 
investigated (e.g., \cite{Greco99,KempR98}).
Let us show that the simpler notion of aggregate stratification 
leads to a unique \CASP\ answer set. The program with 
aggregates  $P$ is aggregate-stratified if there is a
function 
$lev: \Pi_P \mapsto \mathbb{N}$ such that, for 
each rule $H \leftarrow L_1, \dots, L_k$ in $P$,  
\begin{itemize}
\item $lev(pred(H)) \geq lev(pred(L_i))$ if $L_i$ is an ASP-atom; 
\item $lev(pred(H)) > lev(pred(A_i))$ if $L_i$ is the ASP-literal $not\:A_i$; and  
\item $lev(pred(H))> lev(p)$ if $L_i = {\cal P}(s)$ is an aggregate
	atom 
	with $p$ as the predicate of $s$.
\end{itemize}
The notion of  \emph{perfect model} is defined as follows.
\begin{definition}
[Perfect Model, \cite{MumickPR90}]
The \emph{perfect model} of an aggregate-stratified program $P$ is 
the minimal model $M$ such that 
\begin{list}{$\bullet$}{\topsep=-2pt \itemsep=1pt \parsep=0pt}
\item if $M'$ is another model of $P$, then the extension of each predicate
	$p$ of level $0$ in $M$ is a subset of the extension of $p$ in $M'$ 
\item if $M'$ is another model of $P$ such that $M$ and $M'$ agree on the predicates
	of all levels up to $i$, then the extension of each predicate at level
	$i+1$ in $M$ is a subset of the extension of the same predicate in $M'$ 
\end{list}
\end{definition}
       From \cite{MumickPR90,KempS91} we learn that each aggregate-stratified 
program has a unique perfect model. We will show next that 
\CASP-answer sets for aggregate-stratified programs are 
perfect models. 

\begin{theorem}
Let $P$ be an  aggregate-stratified program $P$. The following holds:
\begin{enumerate}
\item If $M$ is an \CASP-answer set of $P$ then $M$ is the 
	perfect model of $P$.
\item The perfect model of $P$ is an \CASP-answer set of $P$.
\end{enumerate}
\end{theorem}
\noindent {\bf Proof.} 
Let $P_i$ be the set of rules in $P$ whose head has the level $i$
and $M(i)$ be the set of atoms in $M$ whose level is $i$. 
\begin{enumerate}
\item 
Let $M$ be an \CASP-answer set of $P$.
Let  $Q = (unfolding(P))^M$. By the definition of answer sets, we  know that 
$M = T_Q\uparrow \omega$ where $T_Q$ is the immediate consequence 
operator for $Q$. 
Since $M$ is an \CASP-answer set of $P$, we know that
$M$ is also a model of $P$ (Theorem \ref{th1}). 
Assume that $M$ is not the perfect model of $P$, i.e.,  the perfect 
model of $P$ is $M'$ and $M \ne M'$.
We have that 
\begin{itemize}
\item $P_0$ is a definite program. Thus,  
	$M(0) = T_{P_0}\uparrow \omega$. This 
	means that $M(0)$ is the least model of $P_0$, which implies
	that $M'(0) = M(0)$.
\item Let us assume that $M$ and $M'$ agree on the levels up to $k$ and
	let us assume $p\in M(k+1)\setminus M'(k+1)$. In particular,
	let us consider the first atom $p$ with such property introduced
	in $M$ by the iterations of $T_Q$. This means that there exists
	a rule $r_2 \in Q$ such that $head(r_2) = p$ and  
	$body(r_2) \subseteq M(0)\cup\dots\cup M(k)$.
	Because $r_2 \in Q$ we can conclude that there exists
	some $r' \in unfolding(P)$ such that $pos(r') = pos(r_2)$ 
	and $neg(r') \cap M = \emptyset$.
	This implies that there exists a rule $r \in P$ 
	and a sequence of aggregate solutions 
	$\langle S_c \rangle_{c \in agg(r)}$ such that 
	$S_c.p \subseteq M$ and $S_c.n \cap M = \emptyset$ 
	for $c \in agg(r)$ and $r'$ is the unfolding of 
	$r$ with respect to 
	$\langle S_c \rangle_{c \in agg(r)}$. Since $M$ and $M'$
	agree on the levels up to $k$, this implies 
	that $M' \models c$ for every $c \in agg(r)$, 
	$pos(r) \subseteq M'$, and $neg(r) \cap M' = \emptyset$.
	Thus, $M' \models body(r)$. Because $M'$ is a model of $P$,
	we have that $p = head(r) = head(r_2) \in M'$. This 
	contradicts the fact that $p \not\in M'$, i.e., 
	$M$ is the perfect model of $P$.
\end{itemize}

\item Let $M$ be the perfect model of $P$. We will show that 
	$M$ is an \CASP-answer set of $P$. 
	Lemma \ref{lem1} implies that
$M$ is a model of $unfolding(P)$, and in particular
$M$ is a model of $Q = (unfolding(P))^M$.
Assume that $M$ is not an \CASP-answer set. This means
that it is not the minimal model of $Q$, i.e., there
exists $M' \subsetneq M$ which is the minimal model of $Q$. We will show
that the existence of $M'$ violates the minimality nature of the perfect
model. 
We have that 
\begin{itemize}
\item $P_0$ is a collection of definite clauses. Thus, $M(0)$ is 
	the least model of $P_0$. Since $M'(0)$ is a model of 
	$P_0$ then we must have $M'(0) = M(0)$.
\item Let $M$ and $M'$ agree on the levels up to $k$; if we consider the 
	program $P_{k+1}$ with the interpretation of all predicates of 
	levels $\leq k$ fixed, we are left with a definite program, whose
	least model is $M(k+1)$ from definition. As $M'(k+1)$ is also a 
	model, we have that it must coincide with $M(k+1)$.
\end{itemize}
This proves the second part of the theorem.	\hfill$\Box$
\end{enumerate}
The following corollary follows directly from the fact that 
an aggregate-stratified program has a unique perfect model
and the above theorem.
\begin{corollary}
Every aggregate-stratified program admits a unique \CASP-answer set.
\end{corollary}
We believe that this equivalence can be easily proved for other forms 
of aggregate-stratification.

\subsection{Monotone Programs}

The notion of \emph{monotone programs} has been introduced in 
\cite{MumickPR90}, and later elaborated by other researchers
(e.g., \cite{KempS91,RossS97}), as another class of programs 
for which the existence of a unique intended model is guaranteed, 
even in presence of recursion 
through aggregation. The notion of monotone programs, 
defined only for programs with aggregates and without negation, is as follows. 

\begin{definition}[Monotone Programs, \cite{KempS91}] \label{monotonic}
Let $F$ be a collection of base predicates and $B$ 
be an interpretation of $F$. A program $P$ is \emph{monotone} 
with respect to $B$ if, for 
each rule $r$ in $ground(P)$ where $pred(head(r)) \notin F$, and for all 
interpretations $I$ and  $I'$, where $B \subseteq I \subseteq I'$, we have that 
$I \models body(r)$ implies $I' \models body(r)$.
\end{definition}
We will follow the convention used in \cite{RossS97} of
 fixing the set of base predicates $F$ to be equal to 
the set of EDB predicates, i.e., it contains only predicates which 
do not occur in the head of rules of $P$. This will also mean that $B$ is fixed
and $B$ is true in every interpretation of the program $P$.  
As such, instead of saying that $P$ is monotone with respect to $B$, 
we will often say that $P$ is monotone whenever there is no confusion.

For a monotone program $P$ with respect to the interpretation $B$ 
of a set of base predicates $F$, the fixpoint operator, denoted by $T_P^B$,
 is extended to include $B$ as follows:
\[
T_P^B(I) = \{head(r) \mid r \in ground(P), \:
			pred(head(r)) \notin F, \: 
                        I \cup B \models body(r)\}. 
\]
It can be shown that $T_P^B$ is monotone and hence has 
a unique least fixpoint, denoted by $lfp(T_P^B)$. We will next 
prove that monotonicity also implies uniqueness of 
\CASP-answer sets. First, we prove a simple observation
characterizing aggregate solutions in monotone programs.

\begin{proposition} \label{obs1}
Let $P$ be a monotone program with respect to $B$ and 
$r$ be a rule in $ground(P)$. Assume that $c \in agg(r)$
and $S_c$ is a solution of $c$. 
Then, $\langle S_c.p, \emptyset \rangle$ is 
also a solution of $c$.
\end{proposition}
\noindent {\bf Proof.} Due to the monotonicity of $P$
we have that $M \models c$ for every interpretation $M$
satisfying the condition $S_c.p \subseteq M$.
This implies that $\langle S_c.p, \emptyset \rangle$ is 
a solution of $c$.
\qed

\begin{theorem}
Let $P'$ be a monotone program w.r.t. $B$ and
let $P = P' \cup B$. Then  $lfp(T^B_{P})$ is an
\CASP-answer set of $P$.
\end{theorem}
\noindent {\bf Proof.}
Let $M= lfp(T_{P}^B)$, $Q =  (unfolding(P))^M$,  and
$M' = T_Q \uparrow \omega$. We will prove that $M = M'$.
First of all, observe that  $B \subseteq M \cap M'$, since
the elements of $B$ are present as facts in $P$. Since the predicates
used in $B$ do not appear as head of any other rule in $P'$, in the
rest we can focus on the elements of $M, M'$ which are distinct from $B$.
\begin{itemize}
\item $M' \subseteq M$: we prove by induction on $k$ that 
$T_Q\uparrow k \subseteq M$. The result is obvious for $k=0$. 
Assume that $T_Q \uparrow k \subseteq M$ and consider 
$p \in T_Q\uparrow k+1$. This
implies that there is a rule $r' \in Q$ 
such that $p = head(r')$ and $pos(r') \subseteq T_Q \uparrow k \subseteq M$.
This means that there exists a  rule $r \in ground(P)$
and a sequence of aggregate solutions $\langle S_c \rangle_{c \in agg(r)}$ 
such that $r'$ is obtained from $r''$, which is the unfolding 
of $r$ with respect to $\langle S_c \rangle_{c \in agg(r)}$,
by removing $neg(r'')$ from its body, i.e., $neg(r'') \cap M = \emptyset$. 
This implies that 
\begin{itemize}
\item $head(r) = head(r')$
\item $pos(r') = pos(r'') = pos(r) \cup \bigcup_{c \in agg(r)} S_c.p$ 
	and $pos(r') \subseteq T_Q \uparrow k \subseteq M$
\item $neg(r'') = neg(r) \cup \bigcup_{c \in agg(r)} S_c.n$ and 
	$neg(r'') \cap M = \emptyset$.
\end{itemize}
This implies that $M \models c$ for every $c\in agg(r)$,
$pos(r) \subseteq M$, and $neg(r) \cap M = \emptyset$. 
This allows us to conclude that $M \models body(r)$.
By the definition of $T_{P'}^B$, we have that $p = head(r) \in M$.

\item $M \subseteq M'$: we will show that 
$T^B_{P}\uparrow k \subseteq M'$ 
for $k \ge 0$. We prove this by induction on $k$. 
The result is obvious for $k=0$.
Assume that $T_P^B\uparrow k \subseteq M'$.
Consider $p \in T_P^B\uparrow k+1$. 
This implies the existence of a rule $r \in ground(P)$ 
such that $head(r) = p$ and $T_P^B\uparrow k \models body(r)$.
This means that 
$pos(r) \subseteq T_P^B\uparrow k \subseteq M'$ 
and $T_P^B\uparrow k \models c$ for every $c \in agg(r)$. From 
Proposition~\ref{obs1}, we know that there exists a sequence 
of aggregate solutions $\langle S_c \rangle_{c \in agg(r)}$ 
such that $S_c.n = \emptyset$ and $S_c.p \subseteq T_P^B\uparrow k$. This 
implies that $r'$, the unfolding of $r$ with respect to 
$\langle S_c \rangle_{c \in agg(r)}$, is a rule in $Q$
and $body(r') \subseteq M'$. Hence, $p = head(r) = head(r') \in M'$.
\end{itemize}
The above results allow us to conclude that $M = M'$.
\hfill$\Box$

\medskip \noindent
Since $lfp(T_P^B)$ is unique, we have the following.
\begin{corollary}
Every monotone program admits exactly one \CASP-answer
set.
\end{corollary}

\subsection{Other Proposals}
\label{old-semantics}
Another semantic characterization of aggregates that has been adopted by 
several researchers \cite{armi-ijcai03,ElkabaniPS04,a-prolog,KempS91}
can be simply described as follows.
Given a program $P$ and an interpretation $M$, let $G(M,P)$ 
be the program obtained by: 
\begin{itemize}
\item[\emph{(i)}] removing all the rules 
with an aggregate atom or a negation-as-failure literal 
which is false in $M$; and 
\item[\emph{(ii)}] removing all the remaining 
aggregate atoms and negation-as-failure literals.
\end{itemize}
 $M$ is a  
{\em stable set} of $P$ if $M$ is the least model of $G(M,P)$.
We can prove the following result.
\begin{theorem}
Let $P$ be a program with aggregates. If $M$ is an \CASP-answer set of
$P$, then $M$ is a stable set of $P$.
\end{theorem}
\noindent{\bf Proof:} 
Let $Q = unfolding(P)^M$ and let us denote with $R=G(M,P)$. Let us show
that $lfp(T_Q) = lfp(T_R)$. 

First, let us show that $lfp(T_Q) \subseteq lfp(T_R)$; we will accomplish
this by showing $T_Q\uparrow k \subseteq lfp(T_R)$ by induction on $k$.
For $k=0$, the result is obvious. Let us assume that $T_Q\uparrow k \subseteq lfp(T_R)$
and let us consider $p \in T_Q \uparrow k+1$. This means that there is a rule
$p \leftarrow pos(r), S.p$ in $Q$ such that $pos(r)\subseteq T_Q\uparrow k\subseteq lfp(T_R)$
and $S.p \subseteq T_Q\uparrow k\subseteq lfp(T_R)$.
This means that there is a rule $p \leftarrow pos(r),\naf neg(r), S.p, \naf S.n$ in
$unfolding(P)$, $M\cap S.n=\emptyset$ and $M\cap neg(r) =\emptyset$. In turn, there
is a rule $p\leftarrow pos(r), \naf neg(r), agg(r)$ in $P$ such that 
$S.p \wedge \naf S.n$ is an unfolding of $agg(r)$. Since $M\cap S.n = \emptyset$
and $S.p \subseteq T_Q\uparrow k \subseteq M$, then $M\models c$. This implies
that $p\leftarrow pos(r)$ is in $R$; since $pos(r) \subseteq lfp(T_R)$ then $p\in lfp(T_R)$.

Second, let us show that $lfp(T_R) \subseteq lfp(T_Q)$; we will accomplish
this by showing that $T_R \uparrow k \subseteq lfp(T_Q)$ by induction on $k$.
The result is obvious for $k=0$. Let us consider $T_R\uparrow k \subseteq lfp(T_Q)$ and
let us now consider $p\in T_R(T_R\uparrow k)$. This means that there is a rule
$p \leftarrow pos(r)$ in $R$ such that $pos(r) \subseteq T_R\uparrow k\subseteq lfp(T_Q)$.
This means that there is a rule $p \leftarrow pos(r), \naf neg(r), agg(r)$ in $P$ such
that $M\models agg(c)$ and $M \cap neg(r) = \emptyset$. This means that there is 
an unfolding of this rule of the form $p \leftarrow pos(r),\naf neg(r), S.p, \naf S.n$
and $S.p \subseteq M$ and $M \cap S.n =\emptyset$. This implies that 
$p\leftarrow pos(r), S.p$ is in $Q$, $pos(r) \subseteq lfp(T_Q)$ and $S.p \subseteq lfp(T_Q)$,
and finally $p \in lfp(T_Q)$.
\hfill$\Box$

\medskip \noindent 
The converse is not true in general, since 
stable sets could be not minimal with respect to set inclusion. For example, 
the program $P_2$ in Example \ref{exp2} has 
$\{p(1),p(2),p(3),p(5),q\}$ as a stable set.

\section{Discussions}
\label{sec-diss}

In this section, we present a program with aggregates in which the 
unfolding transformation (as well as the translation discussed
in \cite{Pelov04}) is not applicable.
We also briefly discuss the computational complexity issues related
to the class of logic programs with aggregates.

\subsection{A Limitation of the Unfolding Transformation}
The key idea of our 
approach lies in that, if an aggregate atom is satisfied in an 
interpretation, one of its solutions must be satisfied. Since our 
main interest is in the class of programs whose answer sets can be 
computed by currently available answer set solvers, we are mainly 
concerned with finite programs and aggregate atoms with finite 
solutions. Here, by a finite solution we mean a solution $S$ whose
components $S.p$ and $S.n$ are finite sets of atoms.
Certain modifications to our approach might be needed 
to deal with programs with infinite domains which can give raise 
to infinite solutions. 
For example, consider the program $P_8$ which consists of 
the rules: 
\[
\begin{array}{lll}
q & \leftarrow & \textnormal{\sc Sum}(X \mid p(X)) \ge 2. \\
p(X/2) & \leftarrow & p(X). \\
p(0).  &  &
p(1). 
\end{array}
\]
It is easy to see that the aggregate atom
$c = \textnormal{\sc Sum}(X \mid p(X)) \ge 2$
has two aggregate solutions, 
$S = \langle Q, \emptyset \rangle$ and 
$T = \langle Q \setminus \{p(0)\}, \emptyset \rangle$,
where 
$
Q = \{p(1/(2^i)) \mid i=0,1,\ldots,\} \cup \{p(0)\}.
$
Both solutions are infinite.
As such, the unfolded version of  program $P_8$ 
is no longer a normal logic program---in the sense that it contains 
some rules whose body is not a {\em finite} set of ASP-literals. 
Presently, it is not clear how the unfolding approach can be
employed in this type of situations.

In \cite{SonP05}, we provide an alternative
definition of \CASP\ 
answer sets which utilizes the notion of solutions but does
not employ the unfolding transformation. This semantics yields
the intuitive answer for $P_8$.  

\subsection{Computational Complexity}

Our main goal in this paper is to develop a framework for dealing 
with aggregates in Answer Set Programming. As we have demonstrated 
in Section \ref{impl}, the proposed semantics can be easily 
integrated to existing answer set solvers. In \cite{SonP05}, we proved 
that the complexity of checking the existence of an answer set 
of a program with aggregates depends on the complexity of 
evaluating aggregate atoms and on the complexity of 
checking aggregate solutions. In particular, we proved that there are
large classes of programs, making use of the standard aggregate
functions (e.g., {\sc Sum}, {\sc Min}), for which the answer set
checking problem is tractable and the problem of determining the
existence of an answer set is in {\bf NP}.
These results are in line with similar results presented in
\cite{Pelov04}.

\section{Conclusions and Future Work}
\label{concl}

In this paper, we presented two equivalent definitions of answer sets 
for logic programs with arbitrary aggregates, and discussed an 
implementation of an answer set solver for programs with aggregates. 
Our definitions are based on a translation process, called {\em unfolding}, 
which reduces programs with aggregates to normal logic programs. The translation builds
on the general idea of \emph{unfolding of intensional sets}
\cite{sets94,neg}, explored in our previous work to handle intensional
sets in constraint logic programming. Key to our definitions is 
the notion of a {\em solution} of an aggregate atom. 

Our first definition can be viewed as an alternative characterization of the
semantics of logic programs with arbitrary  aggregates developed 
in \cite{PelovDB03}. In fact, the first form of unfolding used in  characterizing
the semantics of LP with aggregates
corresponds to an independently developed translation approach proposed
in ~\cite{PelovDB03}, which 
captures the same meaning as the semantics---based
on approximation theory---described in \cite{Pelov04}. 

To allow aggregate atoms in the head, we developed a  second 
translation scheme, which unfolds a program with aggregates w.r.t. a provisional 
answer set. The result of this process is a {\em positive program}
which can be used to verify whether or not the provisional answer set 
is indeed an answer set of the original program. We discussed how 
the second unfolding can be extended to deal with programs with 
aggregate atoms as  heads of rules. 

We discussed the basic components of an implementation based 
on off-the-shelf  answer set solvers, and we described \CASP,
a system capable of computing answer sets of program with 
aggregates. 

We related the semantics for logic programs
with aggregates defined in this paper to other proposals 
in the literature. We showed that it 
coincides with various existing 
proposals on large classes of 
programs (e.g., stratified programs and programs with monotone aggregates).
We also noticed that there are some subtle differences
between distinct semantic characterizations recently proposed
for logic programming with aggregates.

As future work, we propose
to investigate formalizations of semantics of aggregates that can
be parameterized in such a way to cover the most relevant existing
proposals.
Our future work includes also an  investigation of whether 
our alternative characterization for answer sets, based on
unfolding w.r.t. a given interpretation, can  be used to
improve the performance of our implementation.

\subsection*{Acknowledgments}

We would like to thank Vladimir Lifschitz and Michael Gelfond 
for the numerous discussions, 
related to the topics of this paper. We also wish to thank
the anonymous referees of a preliminary version of this work,
for their helpful comments. 
The authors have been supported by 
the NSF grants CNS-0220590, CNS-0454066, and HRD-0420407.
The description of the system \CASP\ has been presented in 
\cite{ElkabaniPS05}.

\bibliographystyle{plain}

\end{document}